\documentclass[fleqn,usenatbib,useAMS]{mnras} 

\usepackage{graphicx}             
\usepackage{amsmath}              
\usepackage{upgreek}              
\usepackage{amssymb}              

\usepackage{bm}                   
\usepackage[mathcal]{euscript}    
\usepackage{CJKutf8}              
\usepackage[T1]{fontenc}          
\usepackage{comment}              
\usepackage{transparent}          
\usepackage{tabularx}             
\usepackage{threeparttable}       
\usepackage{booktabs}             
\usepackage{soul}                 
\usepackage{ulem}                 
\usepackage{multirow}             
\usepackage{makecell}             
\usepackage[dvipsnames]{xcolor}   
\usepackage{tipa}
\usepackage{microtype}            
\DisableLigatures[f]{encoding = *, family = *} 
\usepackage[figuresleft]{rotating}
\emergencystretch=\maxdimen       
\newcommand{\fracp}[2]{\frac{\partial{#1}}{\partial{#2}}}

\newcommand{\hatomega}{\hat{\text{\bf \textscomega}}}

\newcommand{\rlr}[1]{#1}
\newcommand{\biborder}[1]{} 

\defcitealias{Baruteau2011}{B11}
\defcitealias{LiYaPing2021}{L21}
\defcitealias{Li2022}{Paper I}
\defcitealias{Li2023}{Paper II}

\usepackage{ae,aecompl}
\usepackage{multicol}        

\usepackage{newtxtext,newtxmath}

\title[Binaries Embedded in Discs. III]{Hydrodynamical Evolution of Black-Hole Binaries Embedded in AGN Discs: III. The Effects of Viscosity}

\author[R. Li and D. Lai]{
  Rixin Li$^{1,2}$
  \begin{CJK*}{UTF8}{gkai}
    (李日新)
  \end{CJK*} \thanks{Contact e-mail: \href{mailto:rixin@berkeley.edu}{rixin@berkeley.edu}}\thanks{51 Pegasi b Fellow}
  and 
  Dong Lai$^{1,3}$
  \begin{CJK*}{UTF8}{gkai}
    (赖东)
  \end{CJK*}
\\
$^{1}$Center for Astrophysics and Planetary Science, Department of Astronomy, Cornell University, Ithaca, NY 14853, USA \\
$^{2}$Department of Astronomy, University of California Berkeley, Campbell Hall, Berkeley, CA 94720-3411, USA \\
$^{3}$Tsung-Dao Lee Institute, Shanghai Jiao-Tong University, Shanghai, China
}

\date{Last updated 2023 XXX XX; in original form 2023 XXX XX}

\pubyear{2023}

\begin{document}
\label{firstpage}
\pagerange{\pageref{firstpage}--\pageref{lastpage}}
\maketitle

\abovedisplayshortskip=6pt plus 3pt minus 2pt 
\belowdisplayshortskip=6pt plus 3pt minus 2pt 
\abovedisplayskip=6pt plus 3pt minus 2pt 
\belowdisplayskip=6pt plus 3pt minus 2pt 

\begin{abstract}
  Stellar-mass binary black holes (BBHs) embedded in active galactic nucleus (AGN) discs offer a distinct dynamical channel to produce black hole mergers detected in gravitational waves by LIGO/Virgo.  
  To understand their orbital evolution through interactions with the disc gas, we perform a suite of 2D high-resolution, local shearing box, viscous hydrodynamical simulations of equal-mass binaries.
  We find that viscosity not only smooths the flow structure around prograde circular binaries, but also greatly raises their accretion rates.
  \rlr{The torque associated with accretion may be overwhelmingly positive and dominate over the gravitational torque at a high accretion rate.  However, the accreted angular momentum per unit mass decreases with increasing viscosity, making it easier to shrink the binary orbit.
  In addition, }retrograde binaries still experience rapid orbital decay, and prograde eccentric binaries still experience eccentricity damping.
  Our numerical experiments further show that prograde binaries \rlr{are more likely to be hardened}
  if the physical sizes of the accretors are sufficiently small such that \rlr{the accretion rate is reduced.  The dependency}
  of the binary accretion rate on the accretor size can be weaken through boosted accretion either due to a high viscosity or a more isothermal-like equation of state (EOS).  
  Our results widen the explored parameter space for the hydrodynamics of embedded BBHs and demonstrate that their orbital evolution in AGN discs is a complex, multifaceted problem.
\end{abstract}


\section{Introduction}
\label{sec:intro}

The detection of merging binary black holes (BHs) by the LIGO-Virgo-KAGRA collaboration \citep{LIGO_GWTC-3} has stimulated many studies on the dynamical BBH formation channels besides the classical isolated binary evolution channel \citep[e.g.,][]{Lipunov1997, Podsiadlowski2003, Belczynski2010, Belczynski2016}.  These dynamical channels include strong gravitational scatterings in dense star clusters \citep[e.g.,][]{PortegiesZwart2000, Miller2009, Samsing2014, Rodriguez2015, Kremer2019}, more gentle ``tertiary-induced mergers'' (often via Lidov-Kozai mechanism) that take place either in galactic triple/quadrupole systems \citep[e.g.,][]{Miller2002, Silsbee2017, LiuLai2018, LiuLai2019, LiuBin2019b, Fragione2019} or in nuclear clusters dominated by a central supermassive BH \citep[e.g.,][]{Antonini2012, Petrovich2017, Hamers2018, LiuBin2019a, LiuLai2020PRD, LiuLai2021}, and (hydro)dynamical interactions in AGN discs \citep[e.g.,][]{Bartos2017, Stone2017, McKernan2018, Tagawa2020a, Samsing2022}.  The AGN disc channel is of particular interest because it provides deep gravitational potentials where hierarchical mergers are likely \citep[e.g.,][]{Gerosa2021}, and may lead to electromagnetic counterparts to gravitational waves \citep[e.g.,][]{Graham2023}.

In the AGN disc channel, the orbital evolution of the BBHs is initially driven by interactions with the disc gas.  However, whether such interactions lead to orbital decay or expansion is not clear.  \citet{Baruteau2011} and \citet{LiYaPing2021} carried out global simulations of binaries embedded in 2D isothermal viscous discs and found opposite results regarding whether a massive (gap-opening) prograde equal-mass binary would be hardened by dynamical friction.  That said, the former study modelled an non-accreting binary and may not adequately resolve the circum-single disc (CSD) regions.  More works from the latter group \citep{LiYaPing2022, Dempsey2022} further found that the BH feedback on the CSDs, binary separation, and the vertical structure of CSDs may play an important role in the evolution of the binary.

In \citet[][hereafter \citetalias{Li2022}]{Li2022} and \citet[][hereafter \citetalias{Li2023}]{Li2023}, we conducted a series of 2D inviscid hydrodynamical simulations of binaries embedded in AGN discs using local shearing boxes.  We surveyed a range of binary intrinsic properties (i.e., binary eccentricities and mass ratios) and other parameters (i.e., binary semi-major axis relative to its Hill radius, BBH to SMBH mass ratios), and different equation of states (EOSs) for the disc gas.  
We found that prograde comparable-mass circular binaries contract when the EOS is far from isothermal, with an orbital decay rate of a few times the mass doubling rate.  Moreover, we reported that eccentric binaries experience significant eccentricity damping and retrograde binaries yield faster orbital decay.  We further found that the binary orbital evolution considerably depends on the EOS, where prograde binaries instead expand when the EOS is nearly-isothermal. 

A caveat of our prior works --- and also all previous numerical studies with local disc models \citep[e.g.,][]{Dempsey2022} --- is that we only simulated the dynamics of an inviscid flow, while AGN discs are believed to be highly viscous \citep[e.g.,][]{Dittmann2020}.  Previous studies with global disc models did include viscosity, but only a very low viscosity was chosen such that the binary can rapidly open a gap.  Among the global disc works, only \citet{Baruteau2011} investigated the effect of viscosity and found that a higher viscosity would result in a faster binary hardening because the binary bears a larger drag force due to the more massive circumbinary flow.  However, this finding should be taken with caution since gas accretion was not modelled there.

In this paper, we build upon the numerical scheme in \citetalias{Li2022} and, for the first time, take into account viscosity in local box simulations of embedded BBHs.  Our main goal is to understand how viscosity affects the flow structure around BBHs and their long-term orbital evolution.  We are also interested in if viscosity weakens the dependency of the binary accretion rate on the physical size of the accretor. 

The paper is organized as follows.  In Section \ref{sec:methods}, we recapitulate our numerical scheme and detail the calculations of viscosity.  Section \ref{sec:results} presents the simulation results of our main survey, compares the accretion flow morphologies under various setups, and describes how the secular binary orbital evolution varies in the parameter space.  We then report the results of our numerical survey on the sink radius in Section \ref{sec:deps_on_r_s}, followed by a summary of our findings in Section \ref{sec:summary}.

\section{Methods}
\label{sec:methods}

We use the code \texttt{ATHENA} \citep{Stone2008} with a setup similar to \citetalias{Li2022} and \citetalias{Li2023} to study the hydrodynamical evolution of binaries embedded in accretion discs.  Section \ref{subsec:schemes} briefly reiterates our numerical model and describes our treatment of viscosity.  \rlr{Section \ref{subsec:corr_mdot} then describes the updated formulae we use to compute the secular evolution of the
binary.}  Section \ref{subsec:setups} summarizes the parameters used in our simulations.

\subsection{Simulation Setup}
\label{subsec:schemes}

We consider a binary, $m_1$ and $m_2$, centred in a small patch of an accretion disc around a massive object, $M$, using the local shearing box approximation \citep{Goldreich1965, Hawley1995, Stone2010}.  The centre of mass (COM) of the binary is the centre of the computational domain and is located at a fiducial disc radius $R$ from $M$.  At this location, the Keplerian velocity and frequency are $V_{\rm K}=\sqrt{GM/R}$ and $\Omega_{\rm K}=V_{\rm K}/R$, respectively.  Our reference frame rotates at $\Omega_{\rm K}$.

In the rotating frame, we simulate the dynamics of an viscous compressible flow by solving the following equations of gas dynamics in 2D
\begin{align}
  \fracp{\Sigma_{\rm g}}{t} + \nabla \cdot \left(\Sigma_{\rm g} \bm{u}\right) &= {\mathcal{S}_{\rm \Sigma}}, \label{eq:gascon} \\
  \begin{split}\label{eq:gasmom}
    \fracp{(\Sigma_{\rm g} \bm{u})}{t} + \nabla\cdot(\Sigma_{\rm g} \bm{u}\bm{u} + P\bm{I} + \bm{T}_{\rm vis}) &=\\
    \Sigma_{\rm g} \biggl[ 2\bm{u}\times\bm{\Omega}_{\rm K} &+ 2 q_{\rm sh} {\Omega}_{\rm K}^2 \bm{x} - \nabla \phi_{\rm b} \biggr] + {\bm{\mathcal{S}}_{\rm p}}, 
  \end{split} \\
  \begin{split}\label{eq:gasE}
    \fracp{E}{t} + \nabla\cdot\left[ (E + P) \bm{u} + \bm{T}_{\rm vis} \cdot \bm{u} \right] &= \Sigma_{\rm g} \bm{u} \cdot \left(2 q_{\rm sh}\Omega_{\rm K}^2 \bm{x} - \nabla\phi_{\rm b} \right) + {\mathcal{S}_{\rm E}} , \\
    \text{{and}} \ E = \frac{P}{\gamma - 1} &+ \frac{1}{2} \Sigma_{\rm g} (\bm{u} \cdot \bm{u}),
  \end{split}
\end{align}
where $\Sigma_{\rm g}$, $\bm{u}$, $P$, $E$, and $\gamma$ are surface density, velocity, pressure, total energy surface density, and adiabatic index of gas, \rlr{$\mathcal{S}_{\rm \Sigma}$, $\bm{\mathcal{S}}_{\rm p}$, and $\mathcal{S}_{\rm E}$ are the sink terms for mass, momentum, and energy localized in the sink cells}, $\bm{I}$ is the identity matrix, $\bm{T}_{\rm vis}$ is the stress tensor for isotropic viscosity
\begin{equation}
  \bm{T}_{\rm vis} = - 2 \Sigma_{\rm g} \nu \left[\bm{S} - \frac{1}{3} (\nabla\cdot \bm{u}) \bm{I} \right],
\end{equation}
where $\bm{S} = \frac{1}{2} \left[\nabla \bm{u} + (\nabla \bm{u})^{\rm T} \right]$ is the strain-rate tensor, $\nu$ is the isotropic viscous coefficient, $\bm{\Omega}_{\rm K}$ aligns with $\hat{\bm{z}}$, $q_{\rm sh} \equiv \mathbf{-} \textnormal{d}\ln \Omega_{\rm K}/\textnormal{d} \ln R$ is the background shear parameter and is $3/2$ for a Keplerian disc, $\phi_{\rm b}$ is the gravitational potential of the binary
\begin{equation}
    \phi_{\rm b}(\bm{r}_{k}) = -\frac{G m_1}{\sqrt{(\bm{r}_1 - \bm{r}_k)^2 + \xi_{\rm s}^2}} -\frac{G m_2}{\sqrt{(\bm{r}_2 - \bm{r}_k)^2 + \xi_{\rm s}^2}},
\end{equation}
where $\bm{r}_1$ and $\bm{r}_2$ denote the position vectors of the binary components, $\bm{r}_k$ is the centre position of the $k$-th cell in the computational domain, and $\xi_{\rm s}$ is the gravitational softening length.

In this work, we consider both the $\gamma$-law EOS and the isothermal EOS.  For the $\gamma$-law EOS, the sound speed of the gas is given by $c_{\rm s} = \sqrt{\gamma P / \Sigma_{\rm g}}$.  For the isothermal EOS, we do not evolve the energy equation (i.e., Eq. \ref{eq:gasE}) and adopt $P = c_{\rm s}^2 \Sigma_{\rm g}$.  Below we sometimes use $\gamma=1$ to indicate the isothermal EOS for conciseness.  In addition, we use the $\alpha$-disc prescription \citep{Shakura1973}, where $\nu = \alpha c_{\rm s}^2 / \Omega$ with
\begin{equation}
  \Omega \equiv \sqrt{\frac{G m_1}{|\bm{r}_1 - \bm{r}_k|^3} + \frac{G m_2}{|\bm{r}_2 - \bm{r}_k|^3} + \Omega_{\rm K}^2}.
\end{equation}

The binary in our models has total mass $m_{\rm b} = m_1 + m_2$ and orbits on a prescribed orbit with a semi-major axis of $a_{\rm b}$.  The mean orbital frequency, orbital angular momentum, and energy in the inertial frame are thus
\begin{align}
  \bm{\Omega}_{\rm b} &= \sqrt{\frac{G m_{\rm b}}{a_{\rm b}^3}}\ \hatomega_{\rm b} = \frac{v_{\rm b}}{a_{\rm b}} \hatomega_{\rm b}, \qquad \text{with}~ v_{\rm b} \equiv \sqrt{\frac{G m_{\rm b}}{a_{\rm b}}}, \\
  \bm{L}_{\rm b} &= \mu_{\rm b} \bm{\ell}_{\rm b} = \mu_{\rm b} \bm{\Omega}_{\rm b} a_{\rm b}^2, \\
  E_{\rm b} &= \mu_{\rm b} \mathcal{E}_{\rm b} = -\mu_{\rm b} \frac{G m_{\rm b}}{2 a_{\rm b}},
\end{align}
where $\hatomega_{\rm b}$ is binary normal unit vector, $\mu_{\rm b} = m_1 m_2 / m_{\rm b}$ is the reduced mass, and $\ell_{\rm b}$ and $\mathcal{E}_{\rm b}$ are the specific angular momentum and specific energy, respectively.  In the rotating shearing box frame, the apparent binary orbital frequency (or period) is $\bm{\Omega}_{\rm b}'$ (or $P_{\rm b}'$).  Throughout this paper, we consider co-planar equal-mass binaries on prescribed orbits (see Section 2.1 of \citetalias{Li2022} for detailed prescriptions).

The binary interacts with the background flow, which is given by
\begin{equation}\label{eq:v_wind}
  \begin{aligned}
    \bm{V}_{\rm w} &= \bm{V}_{\rm sh} + \bm{\Delta V}_{\rm K} \\
    &= -\frac{3}{2} \Omega_{\rm K} x \bm{\hat{y}} - \beta h^2 V_{\rm K} \bm{\hat{y}}
  \end{aligned}
\end{equation}
in the shearing box frame, where $\bm{V}_{\rm sh}(x)$ denotes the Keplerian shear, $\bm{\Delta V}_{\rm K}$ is the deviation from Keplerian velocity, $h\equiv H_{\rm g}/R = c_{\rm s,\infty} / \Omega_{\rm K} R$ is the disc aspect ratio (with $H_{\rm g}$ the disc scale height and $c_{\rm s,\infty}$ the sound speed of the background gas), and $\beta \simeq - \textnormal{d}\ln P / \textnormal{d}\ln R $ is an order unity coefficient determined by the (background) disc pressure profile and is fixed to $1$ in this work.

Besides $\gamma$ and $\alpha$, we expect that our results depend on two dimensionless parameters: $q/h^3{\equiv (m_{\rm b}/M)(R/H_{\rm g})^3 \equiv (R_{\rm H}/H_{\rm g})^3}$ and $\lambda \equiv R_{\rm H}/a_{\rm b}$, where \rlr{$q=m_{\rm b}/M$ is the mass ratio of the binary to the massive object in the disc center,} $R_{\rm H} = R (m_b/M)^{1/3}$ is the Hill radius.  The other dimensionless ratios are all related to $q/h^3$ and $\lambda$:
\begin{align}
  \frac{R_{\rm H}}{H_{\rm g}} &= \left( \frac{R_{\rm B}}{H_{\rm g}} \right)^{1/3} = \left( \frac{q}{h^3} \right)^{1/3}, \label{eq:R_H_over_H} \\
  \frac{v_{\rm b}}{c_{\rm s,\infty}} &= \lambda^{1/2} \left(\frac{q}{h^3} \right)^{1/3}, \\
  \frac{V_{\rm s}}{c_{\rm s,\infty}} &= \frac{|V_{\rm sh}(x=a_{\rm b})|}{c_{\rm s,\infty}} = \frac{3}{2\lambda} \left(\frac{q}{h^3} \right)^{1/3},
\end{align}
where $R_{\rm B} = G m_{\rm b} / c_{\rm s,\infty}^2$ is the Bondi radius.

\begin{figure*}
  \centering  
  \includegraphics[width=\linewidth]{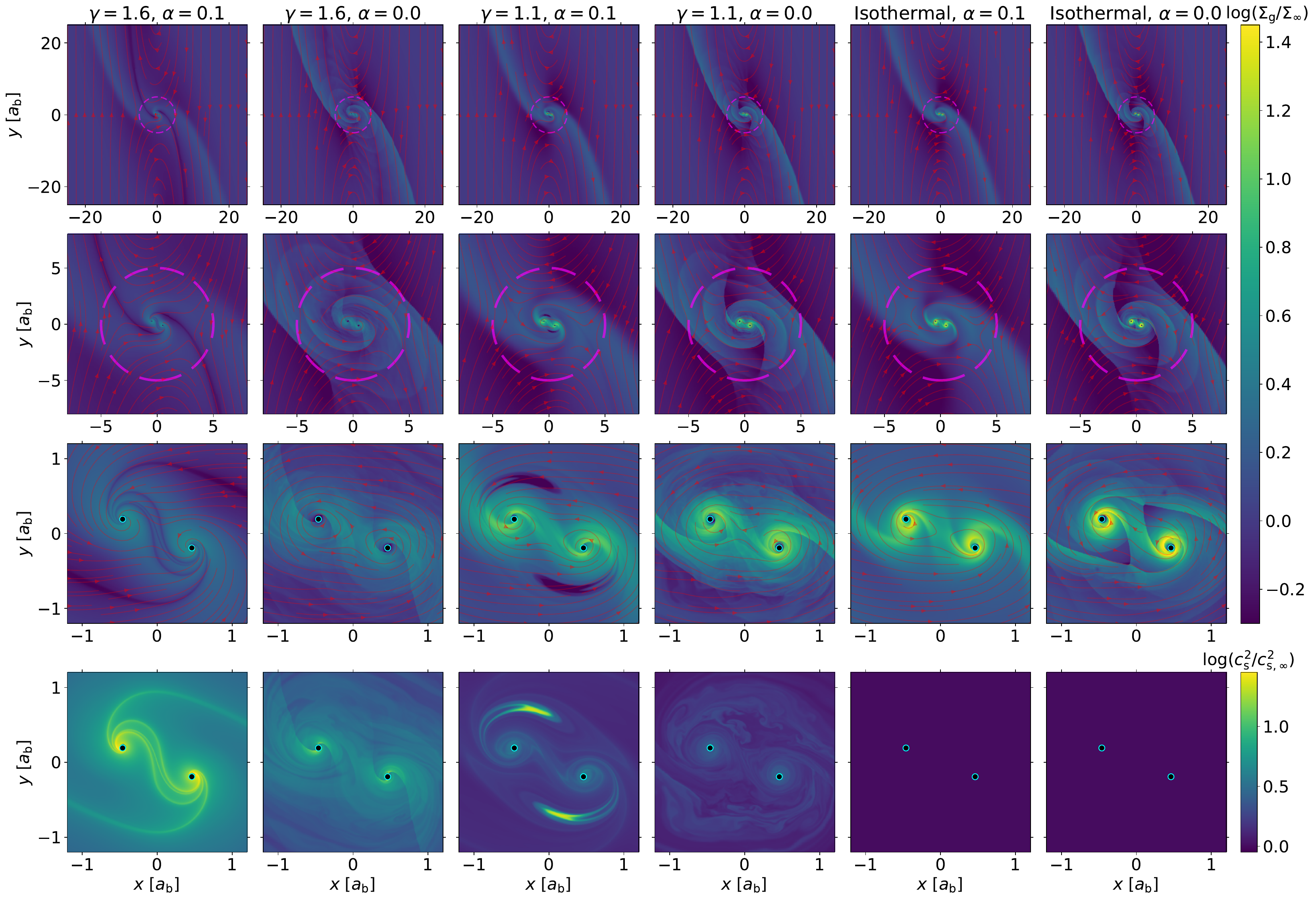}
  \caption{Final quasi-steady snapshots for runs with $\lambda=5$, $q/h^3=1$, $\gamma=1.6$, $1.1$, and isothermal (from \textit{left} to \textit{right}), and $\alpha=0.1$ and $0.0$ (in \textit{alternating rows}), where the mesh is refined progressively towards the binary (zooming in from \textit{top} to \textit{bottom}), showing the gas surface density and detailed flow structures (\textit{red} streamlines) in the first three rows and the thermal structures in the bottom row.  In the \textit{upper} two rows, the \textit{magenta dashed} circles show the Hill sphere ($R_{\rm H}$).  \rlr{In the \textit{lower} two rows, the \textit{cyan solid} circles represent the sink radius ($r_{\rm s}$) of each binary component.}
  \label{fig:snap_runII_qth1}}
\end{figure*}

\begin{figure*}
  \centering
  \includegraphics[width=\linewidth]{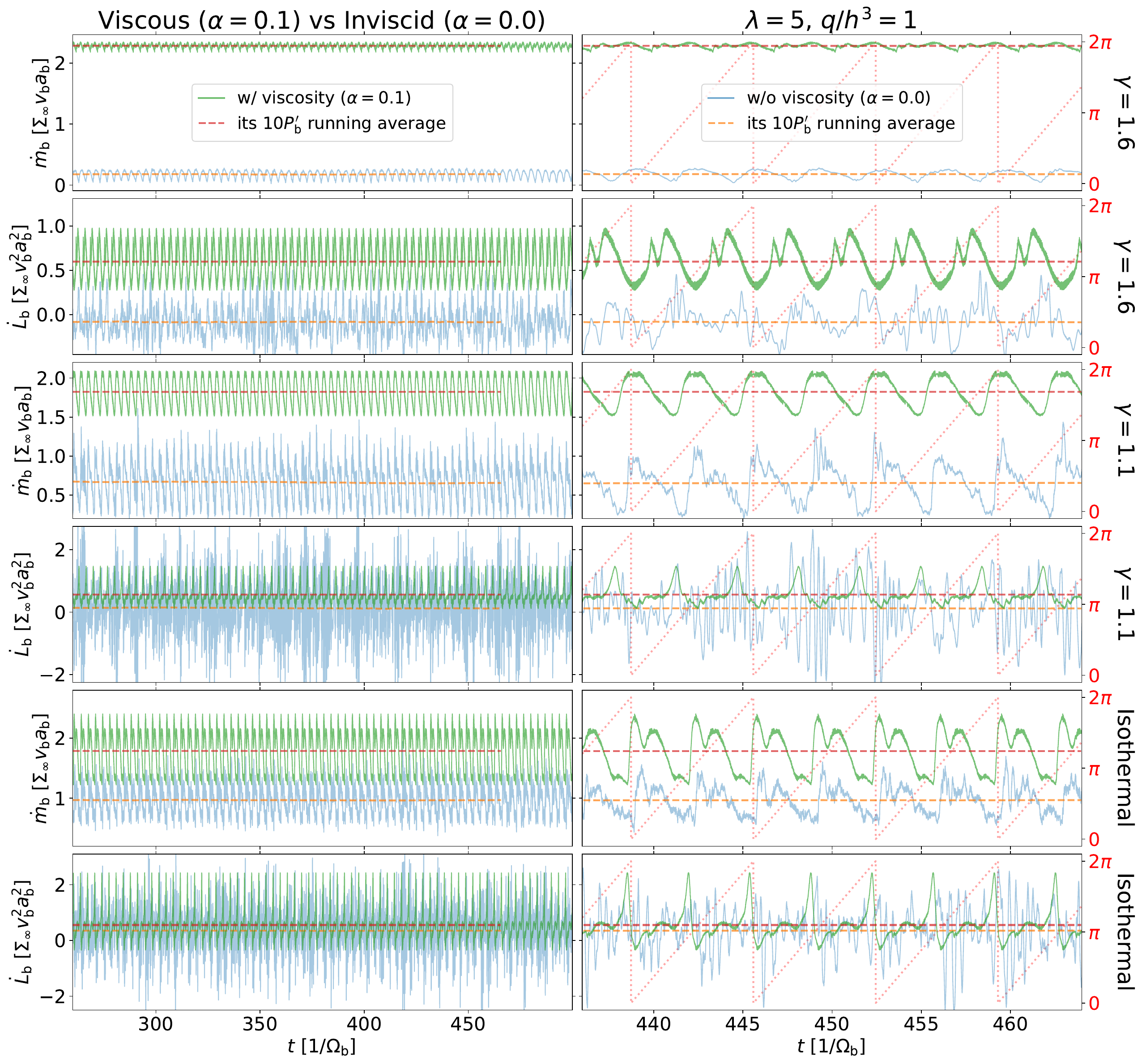}
  \caption{Comparisons of the time series of accretion rate $\dot{m}_{\rm b}$ and total torques $\dot{L}_{\rm b}$ on the binary in the viscous runs ($\alpha=0.1$; \textit{green}) and inviscid runs ($\alpha=0.0$; \textit{blue}) with $q/h^3=1$ and with $\gamma=1.6$, $1.1$, and isothermal (from \textit{top} to \textit{bottom}), for the whole time used for time-averaging (\textit{left}) and a small slice of time (\textit{right}).  Each time series is accompanied by the $10P_{\rm b}'$ running average (\textit{dashed red} for viscous runs and \textit{dashed orange} for inviscid runs). In the right panels, each slice of time series is accompanied by the binary orbital phase curve with period $P_{\rm b}'$ (\textit{dotted pink}).
  \label{fig:dot_mL_runII_qth1}}
\end{figure*}

\begin{figure*}
  \centering
  \includegraphics[width=0.495\linewidth]{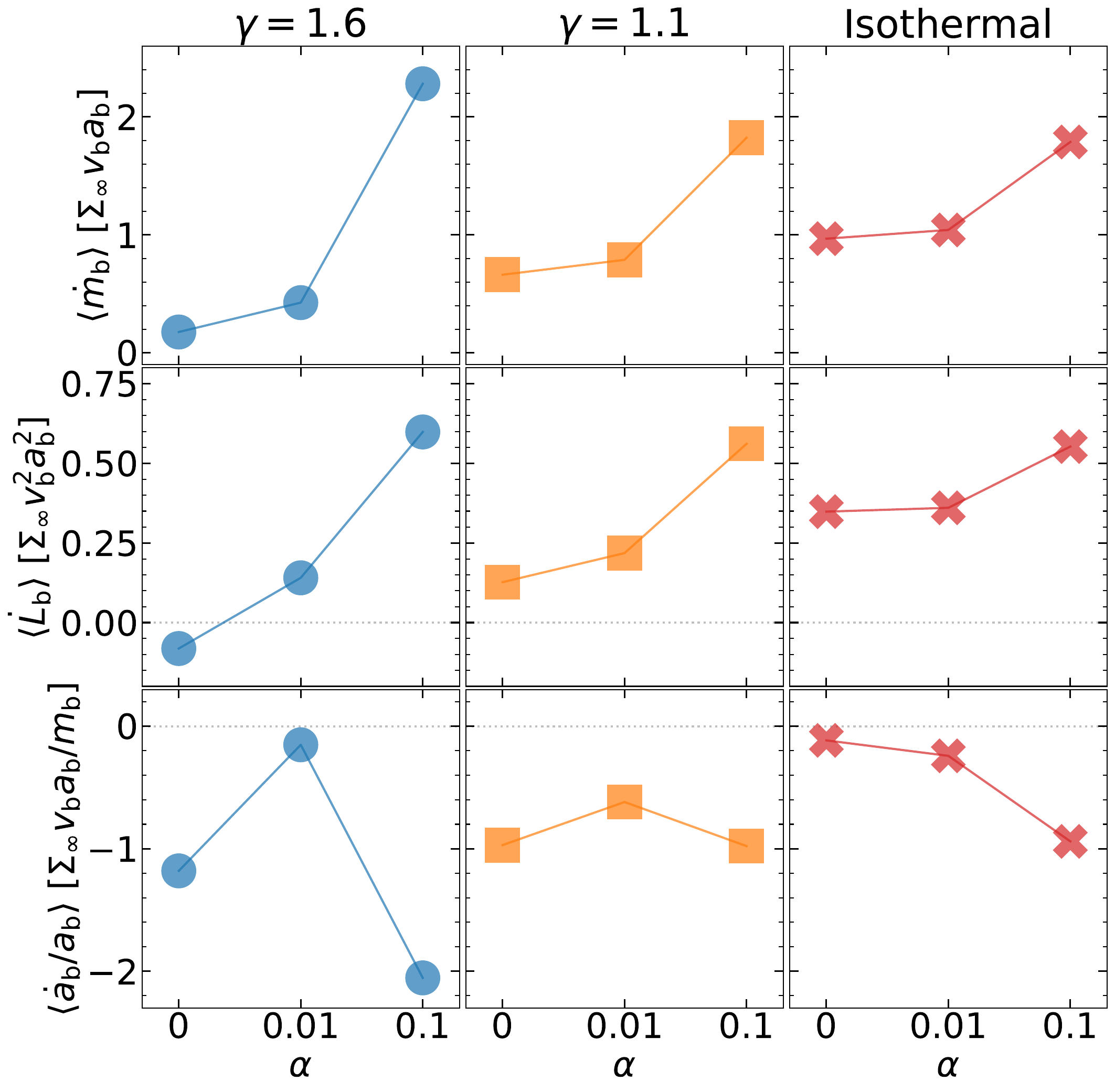}
  \includegraphics[width=0.495\linewidth]{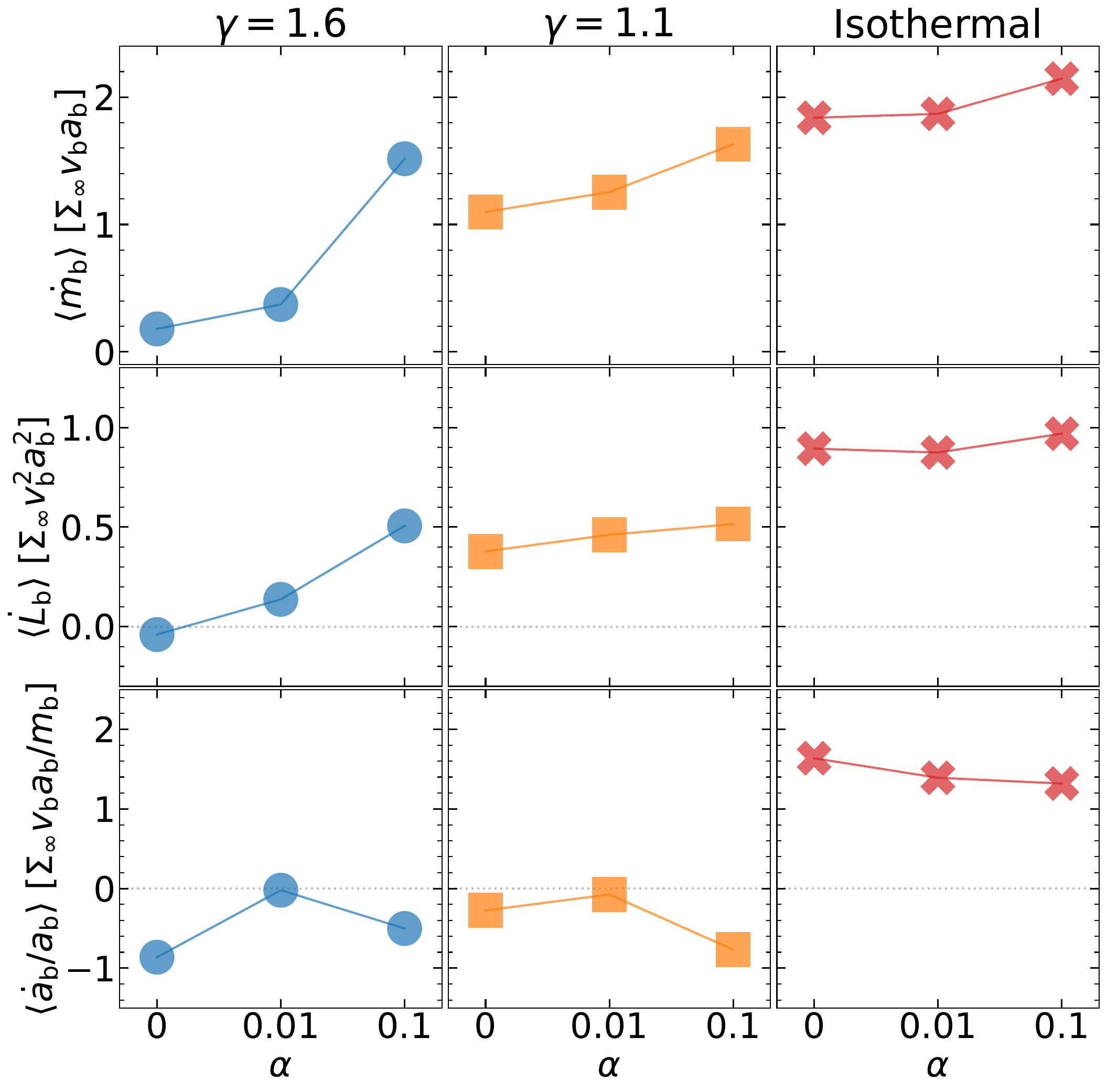}
  \caption{Time-averaged measurements of (from \textit{top} to \textit{bottom}) accretion rate $\langle\dot{m}_{\rm b}\rangle$, total torque $\langle\dot{L}_{\rm b}\rangle$, and binary migration rate $\langle\dot{a}_{\rm b}\rangle/a_{\rm b}$ as a function of $\alpha$, color-coded by the EOS ($\gamma=1.6$: \textit{blue circle}; $1.1$: \textit{orange square}; isothermal: \textit{red cross}), for runs with $q/h^3=1$ (\textit{left}) and with $q/h^3=3$ (\textit{right}).
  \label{fig:runII_trends_alpha}}
\end{figure*}

\subsection{\rlr{Calculations of Accretion Rate, Torques, and Orbital Evolution}}
\label{subsec:corr_mdot}

\rlr{In a similar manner as in \citetalias{Li2022} and \citetalias{Li2023}, we model each binary component as an absorbing sphere with a sink radius of $r_{\rm s}$ and measure the accretion rates and torques on the binary on-the-fly in each time-step in order to determine the time-averaged long-term binary orbital evolution (see Section 2.2 in \citetalias{Li2022}).  A new force term and thus a new torque term is added in this work to account for the inclusion of viscosity.}

\rlr{Specifically, we linearly interpolate the conservative variables of \texttt{ATHENA} onto structured polar grid points around each accretor at an evaluation radius $r_{\rm e}$ in each time step, where we perform the following integrations to obtain the accretion rate and the specific force due to accretion, pressure, and viscosity}
\begin{align}
  \dot{m}_i &= \oint d\dot{m}_i = \oint \left[ -\Sigma_{\rm g} (\bm{u} - \bm{v}_{i,\mathrm{SB}}) \right] \cdot \textnormal{d}\bm{A}, \label{eq:m_dot_i} \\
  \bm{f}_{\mathrm{acc},i} &= \frac{1}{m_i} \oint \textnormal{d}\dot{m}_i (\bm{u} - \bm{v}_{i,\mathrm{SB}}), \label{eq:f_hydro}\\
  \bm{f}_{\mathrm{pres},i} &= \frac{1}{m_i} \oint (-P) \ \textnormal{d}\bm{A}, \label{eq:f_pres} \\
  \bm{f}_{\mathrm{visc},i} &= \frac{1}{m_i} \oint (-T_{\rm vis}) \ \textnormal{d}\bm{A}, \label{eq:f_visc}
\end{align}
\rlr{where $\textnormal{d}\bm{A}$ is the area (line) element around each accretor, $\bm{v}_{i,\mathrm{SB}} = \bm{v}_i + \bm{\Omega}_{\rm pre}\times \bm{r}_i$ is its velocity in the shearing box reference frame.  Note that Eq. \ref{eq:m_dot_i} above differs from Eq. 22 in \citetalias{Li2022}, which used $(-\Sigma_{\rm g} \bm{u}) \cdot \textnormal{d}\bm{A}$ for integration.  It is more appropriate to use the relative velocity between the moving accretor and the surrounding gas to account for the mass flux.  That said, given that the gas is highly super-sonic (i.e., $u \gg c_{\rm s,\infty}$) and moves faster than the accretor, the time-averaged accretion rate $\langle \dot{m}_{\rm b} \rangle$ from Eq. \ref{eq:m_dot_i} and the accretion force/torque are expected to only differ modestly from those obtained with the previous method.  Appendix \ref{appsec:corrected_mdot} demonstrates that the secular orbital evolution results for our previous simulation surveys only change slightly and all the trends and parameter dependencies identified previously still hold.}

\rlr{The net hydrodynamical force (per unit mass)
from the gas on each binary component is then}
\begin{equation}
  \bm{f}_i \equiv \bm{f}_{\mathrm{acc},i} + \bm{f}_{\mathrm{pres},i} + \bm{f}_{\mathrm{grav}, i} + \bm{f}_{\mathrm{visc},i},
\end{equation}
\rlr{where $\bm{f}_{\mathrm{grav}, i}$ is determined in the same way as Section 2.2 in \citetalias{Li2022}.  The resulting total torque that governs the binary orbital evolution becomes}
\begin{equation}
  \dot{\bm{L}}_{\rm b} \equiv \dot{\bm{L}}_{\rm b,acc} + \dot{\bm{L}}_{\rm b,pres} + \dot{\bm{L}}_{\rm b,grav} + \dot{\bm{L}}_{\rm b,visc}, \label{eq:dotL_breakdown}
\end{equation}
\rlr{where}
\begin{align}
  \dot{\bm{L}}_{\rm b,acc} &\equiv \mu_{\rm b} \bm{r}_{\rm b} \times (\bm{f}_{\mathrm{acc},1} - \bm{f}_{\mathrm{acc},2}) + \dot{\mu}_{\rm b} (\bm{r}_{\rm b} \times \dot{\bm{r}}_{\rm b}), \label{eq:dotL_acc} \\
  \dot{\bm{L}}_{\rm b,pres} &\equiv \mu_{\rm b} \bm{r}_{\rm b} \times (\bm{f}_{\mathrm{pres},1} - \bm{f}_{\mathrm{pres},2}), \label{eq:dotL_pres} \\
  \dot{\bm{L}}_{\rm b,grav} &\equiv \mu_{\rm b} \bm{r}_{\rm b} \times (\bm{f}_{\mathrm{grav},1} - \bm{f}_{\mathrm{grav},2}). \label{eq:dotL_grav} \\
  \dot{\bm{L}}_{\rm b,visc} &\equiv \mu_{\rm b} \bm{r}_{\rm b} \times (\bm{f}_{\mathrm{visc},1} - \bm{f}_{\mathrm{visc},2}), \label{eq:dotL_visc}
\end{align}
\rlr{where Eqs. \ref{eq:dotL_acc}-\ref{eq:dotL_grav} are the same as Eqs. 30-32 in \citetalias{Li2022}.  With these updated specific force $f_i$ and total torque $\dot{\bm{L}}_{\rm b}$, we follow Eqs. 33-41 in \citetalias{Li2022} to calculate the secular binary orbital evolution, i.e., the secular rates of change in $a_{\rm b}$ and $e_{\rm b}$.}

\rlr{Furthermore, in this work we also evaluate the spin torque, i.e., the amount of angular momentum transferred to each component of the binary,}
\begin{equation}
  \dot{\rm{S}}_i = \oint \left[ (\bm{r}_{\textnormal{d}A} - \bm{r}_i) \times \Sigma_{\rm g} (\bm{u} - \bm{v}_{i,\mathrm{SB}}) \right] \textnormal{d}A, \label{eq:S_dot_i}
\end{equation}
\rlr{where $\bm{r}_{\textnormal{d}A}$ is the position vector of the area (line) element around each accretor.}

\begingroup 
\setlength{\medmuskip}{0mu} 
\setlength\tabcolsep{4pt} 
\setcellgapes{3pt} 
\begin{table}
  \nomakegapedcells
  \caption{Results for the main simulation survey with varying viscosity} \label{tab:runs-visc}
  \makegapedcells 
  \linespread{1.025} 
  \begin{tabular}{ccc|rrrr}
    \hline
    $q/h^3$
    & $\gamma$
    & $\alpha$
    & \makecell[c]{$\langle\dot{m}_{\rm b}\rangle$}
    & \makecell[c]{$\langle\dot{L}_{\rm b}\rangle$}
    & \makecell[c]{$\displaystyle \frac{\langle\dot{a}_{\rm b}\rangle}{a_{\rm b}}$}
    & \makecell[c]{$\displaystyle \frac{ \left\langle\dot{S}_i \right\rangle }{ \langle\dot{m}_i \rangle }$}
    \\
    
    &
    & 
    & \makecell[c]{\footnotesize $\displaystyle \left[ \Sigma_\infty v_{\rm b} a_{\rm b} \right]$}
    & \makecell[c]{\footnotesize $\displaystyle \left[ \Sigma_\infty v_{\rm b}^2 a_{\rm b}^2 \right]$}
    & \makecell[c]{\small $\displaystyle \left[ \frac{\Sigma_\infty a_{\rm b} v_{\rm b}}{m_{\rm b}} \right]$}
    & \makecell[c]{\small $\displaystyle \left[ \sqrt{G m_i r_{\rm e}} \right]$}
    \\
    (1)
    & (2)
    & (3)
    & \makecell[c]{(4)}
    & \makecell[c]{(5)}
    & \makecell[c]{(6)}
    & \makecell[c]{(7)}
    \\

    \hline\hline
       &      &  $0.0$ &$ 0.18$ &$-0.08$ &$-1.18$ &$ 1.08$ \\
       & $1.6$& $0.01$ &$ 0.43$ &$ 0.14$ &$-0.15$ &$ 1.17$ \\
       &      &  $0.1$ &$ 2.28$ &$ 0.60$ &$-2.05$ &$ 0.78$ \\ \cline{2-7}
       &      &  $0.0$ &$ 0.66$ &$ 0.13$ &$-0.97$ &$ 1.06$ \\
    $1$& $1.1$& $0.01$ &$ 0.79$ &$ 0.22$ &$-0.62$ &$ 1.05$ \\
       &      &  $0.1$ &$ 1.82$ &$ 0.56$ &$-0.98$ &$ 1.00$ \\ \cline{2-7}
       &      &  $0.0$ &$ 0.97$ &$ 0.35$ &$-0.12$ &$ 1.06$ \\
       &   $1$& $0.01$ &$ 1.04$ &$ 0.36$ &$-0.24$ &$ 1.05$ \\
       &      &  $0.1$ &$ 1.79$ &$ 0.55$ &$-0.94$ &$ 1.03$ \\
    \Xhline{3\arrayrulewidth}
       &      &  $0.0$ &$ 0.18$ &$-0.04$ &$-0.86$ &$ 1.10$ \\
       & $1.6$& $0.01$ &$ 0.37$ &$ 0.14$ &$-0.02$ &$ 1.16$ \\
       &      &  $0.1$ &$ 1.52$ &$ 0.51$ &$-0.50$ &$ 0.80$ \\ \cline{2-7}
       &      &  $0.0$ &$ 1.10$ &$ 0.38$ &$-0.27$ &$ 1.05$ \\
    $3$& $1.1$& $0.01$ &$ 1.25$ &$ 0.46$ &$-0.07$ &$ 1.04$ \\
       &      &  $0.1$ &$ 1.63$ &$ 0.52$ &$-0.77$ &$ 1.00$ \\ \cline{2-7}
       &      &  $0.0$ &$ 1.84$ &$ 0.89$ &$ 1.64$ &$ 1.06$ \\
       &   $1$& $0.01$ &$ 1.87$ &$ 0.87$ &$ 1.39$ &$ 1.05$ \\
       &      &  $0.1$ &$ 2.14$ &$ 0.97$ &$ 1.32$ &$ 1.05$ \\
    \Xhline{3\arrayrulewidth}
    $3^{\dagger}$ & $1.1$&  $0.1$&$ 3.72$ &$ 3.09$ &$-35.90$ &$ 0.05$ \\
    $3^{\ddagger}$& $1.1$&  $0.1$&$ 1.97$ &$ 0.88$ &$ -1.33$ &$ 1.06$ \\
    \hline
  \end{tabular} \\
  \begin{flushleft}
    {\large N}OTE -- Columns:
    (1) $q/h^3 = (m_{\rm b}/M)(H_{\rm g}/R)^{-3} = (R_{\rm H}/H_{\rm g})^3$;
    (2) $\gamma$ in the $\gamma$-law EOS (the cases with $\gamma=1$ are isothermal runs);
    (3) $\alpha$-viscosity; 
    (4) time-averaged accretion rate;
    (5) time-averaged rate of change of the binary angular momentum;
    (6) binary semimajor axis change rate;
    (7) ratio between time-averaged spin torque and time-average accretion rate, averaged across the two binary components ($\left\langle\dot{S}_1 \right\rangle / \langle\dot{m}_1 \rangle$ and $\left\langle\dot{S}_2 \right\rangle / \langle\dot{m}_2 \rangle$ are almost identical due to symmetry).
    \\
    $^{\dagger}$ -- This run models a retrograde circular binary.  
    \\
    $^{\ddagger}$ -- This run models a prograde eccentric binary with $e_{\rm b}=0.5$, where the measured binary eccentricity change rate is $\langle \dot{e}_{\rm b}^2 \rangle = -1.36 \langle\dot{m}_{\rm b}\rangle/m_{\rm b}$. \\
  \end{flushleft}
\end{table}
\endgroup

\begin{figure*}
  \centering
  \includegraphics[width=0.495\linewidth]{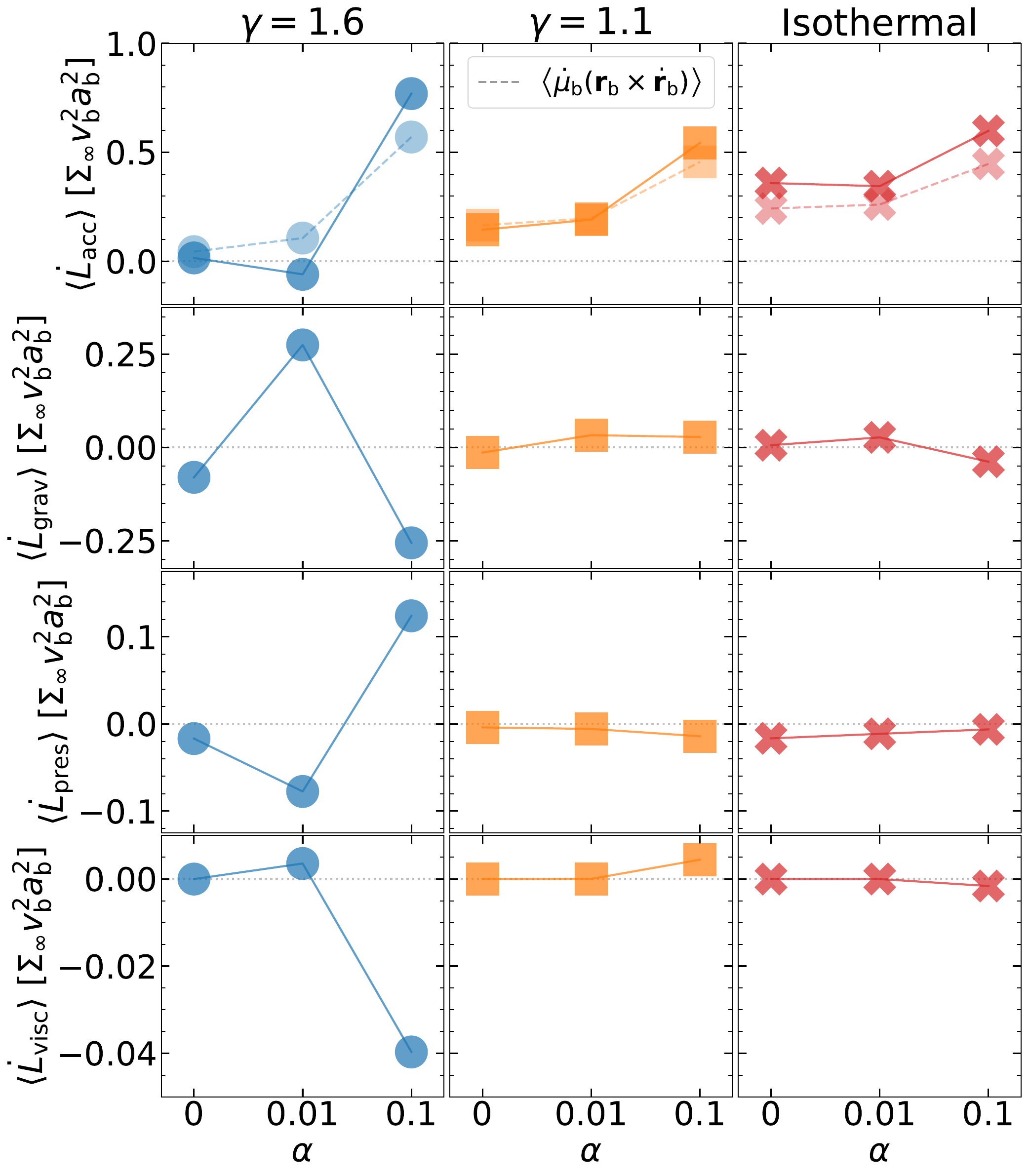}
  \includegraphics[width=0.495\linewidth]{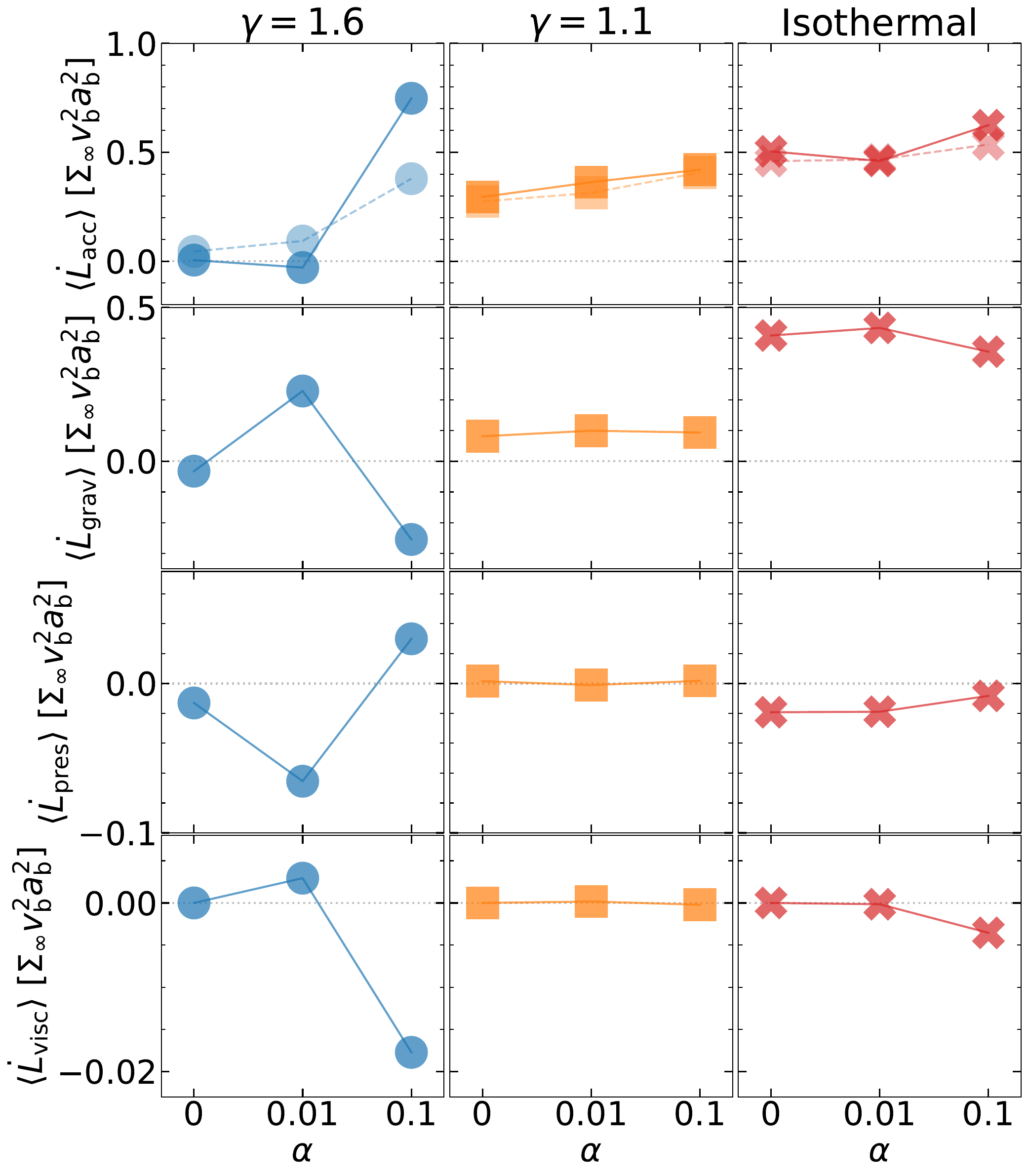}
  \caption{Similar to Fig. \ref{fig:runII_trends_alpha} but showing the time-averaged (from \textit{top} to \textit{bottom}) decomposed torques, $\langle\dot{L}_{\rm acc/grav/pres/visc}\rangle$ \rlr{as a function of $\alpha$}, for runs with $q/h^3=1$ \rlr{(\textit{left}) and with $q/h^3=3$ (\textit{right})}.  \rlr{The top row further shows the contribution (\textit{dashed} lines) from the second term in Eq. \ref{eq:dotL_acc}.}
  \label{fig:runII_dotLs_trends_alpha}}
\end{figure*}

\subsection{Numerical Parameters}
\label{subsec:setups}

The flow dynamics and results of our simulations depend on the dimensionless parameters ($q/h^3$, $\lambda$), the EOS ($\gamma$), and the viscosity ($\alpha$).  Tables \ref{tab:runs-visc} summarizes the parameters used in our main set of runs for prograde \rlr{equal-mass circular} binaries, where we fix $\lambda=5$ and survey $q/h^3=1$ and $3$, a range of $\gamma$ ($1.6$, $1.1$, and isothermal)
\footnote{Compared to \citetalias{Li2023}, we drop $\gamma=1.001$ since the cases with $\gamma=1.001$ showed excellent agreement with the isothermal cases.}
, and a series of $\alpha$ ($0.0$, $0.01$, $0.1$).  In addition, we perform two experiments with $(q/h^3, \gamma, \alpha)=(3, 1.1, 0.1)$ --- one with a retrograde circular binary and the other with a prograde eccentric binary with $e_{\rm b}=0.5$.

Similar to \citetalias{Li2022} and \citetalias{Li2023}, the gas in all of our simulations are initialized with $\Sigma_{\rm g} = \Sigma_{\infty}$ and with the velocity given by the background wind profile $\bm{V}_{\rm w}$ (see Eq. \ref{eq:v_wind}).  We set the root computational domain size to $10 R_{\rm H}$ in both $x$ and $y$ directions.  At all boundaries of the root domain, we adopt a wave-damping open boundary condition that damps the flow back to its initial state with a wave-damping timescale $P_{\rm d} = 0.02\Omega_{\rm b}^{-1}$.  To properly resolve the flow around the binary, we employ multiple levels of static mesh refinement (SMR) towards the binary, where the resolution at the finest level is $a_{\rm b}/\delta_{\rm fl} = 245.76$, where $\delta_{\rm fl}$ is the cell size at that level.  In all simulations in our main survey, we adopt a sink radius of $r_{\rm s}=0.04 a_{\rm b}$, an evaluation radius of $r_{\rm e}=r_{\rm s}+\sqrt{2} \delta_{\rm fl}$, and a gravitational softening length of $10^{-8} a_{\rm b}$.

In each simulation, we prescribe the binary orbital motion and evolve the flow dynamics for $500\Omega_{\rm b}^{-1}$.  With the accretion rates and torques measured on-the-fly in each time-step, the long-term binary orbital evolution is determined by the time-averaged long-term measurements in the post-processing analyses (see Section 2.2 in \citetalias{Li2022} \rlr{and Section \ref{subsec:corr_mdot} above}).  The time-averaging is done over the last $240\Omega_{\rm b}^{-1}$.

\section{Results}
\label{sec:results}

Tables \ref{tab:runs-visc} summarizes the key parameters and results of our simulation suite in the three dimensional parameter space of $q/h^3$, $\gamma$, and $\alpha$.  We first describe the accretion flow morphologies in Section \ref{subsec:flow_field}, and present our findings on the secular orbital evolution in Section \ref{subsec:orbital_evolution}, followed by case studies on a retrograde binary and on an eccentric binary in Section \ref{subsec:retrograde_eccentric}.

\subsection{Flow Structure}
\label{subsec:flow_field}

Fig. \ref{fig:snap_runII_qth1} compares the flow structure between the viscous runs ($\alpha=0.1$) and the inviscid runs ($\alpha=0.0$) in the final quasi-steady state.  We find that strong viscosity diffuses and thus smooths sharp shock fronts.  All the shocks seen in the inviscid runs become suppressed or even imperceptible in the viscous runs, including the grand half bow shocks extending from the Hill sphere to the $\pm y$ boundaries, the large spiral shocks winding from the binary towards the Hill sphere, the waves of prior large spiral shocks propagating along the grand half bow shocks, and the small spiral shocks originated from each binary component.

Furthermore, since strong viscosity greatly facilitates gas accretion onto the binary, weakening the need for the CSDs to act as an accretion buffer, we find that the CSDs in the viscous runs are moderately less massive than their counterparts in the inviscid runs.  When the EOS is far from isothermal (i.e., the $\gamma=1.6$ cases), there are no persistent CSDs because the binary is able to directly accrete the ambient gas that is hot and diffuse (see \citetalias{Li2023} for details on how the CSDs depend on the EOS).  Moreover, since the binary accretes much more efficiently, a pair of corridors with low surface density ($\Sigma_{\rm g}$) but high temperature (and thus $c_{\rm s}^2$ and $\nu$) appear.  They originate from the binary components and then spiral outward, separating the horseshoe flows and the shear flows that are flowing away from the binary, suggesting that certain streamlines flowing towards the binary are directly absorbed by them.  There are also a few remnant corridors directly connecting the two binary components, which resulted from periodic interceptions between one binary component and the original corridor stemming from the other component during orbital motion.  Such corridors are not seen in the inviscid run with $\gamma=1.6$, where there is merely a weak shock front between the two flow regions.

In the viscous run with $\gamma=1.1$, we find that a pair of crescent voids (relatively depleted in $\Sigma_{\rm g}$ but with high $c_{\rm s}^2$ and $\nu$) form in the circumbinary flow, with each of them trailing one binary component.  This feature is again due to the faster accretion onto the binary such that gas in certain regions cannot be replenished in time\footnote{A video of this simulation is available at \url{https://youtu.be/5za0cp7jT38}, which shows the gradual formation of the crescent voids.}.

\subsection{Secular Evolution of Binary}
\label{subsec:orbital_evolution}

Using the post-processing procedure described in \citetalias{Li2022} \rlr{and in Section \ref{subsec:corr_mdot}}, we compute the time series of accretion rate and torques onto the binary in each simulation.  Fig. \ref{fig:dot_mL_runII_qth1} compares the time series of $\dot{m}_{\rm b}$ and $\dot{L}_{\rm b}$ obtained in the viscous runs and inviscid runs with $q/h^3=1$.  We find that the time series curves from the viscous runs are much smoother than those from the inviscid runs, which is consistent with our finding in Section \ref{subsec:flow_field} that shocks are damped by viscosity.  The lack of stochastic small-scale turbulence due to shocks largely reduces the high-frequency modulations in the viscous time series data, which further leads to periodic variations with a coherent and steady amplitude at the binary orbital frequency (${2\Omega_{\rm b}'}$ to be exact).

Fig. \ref{fig:runII_trends_alpha} shows the secular orbital evolution results as a function of $\alpha$ for our survey with different $q/h^3$ and $\gamma$.  We find that viscosity boosts the accretion rate, particularly when the EOS is far from isothermal (see Fig. \ref{fig:dot_mL_runII_qth1}; see also \citetalias{Li2023} for details on how $\langle\dot{m}_{\rm b}\rangle$ depends on the EOS in the inviscid cases).  For the runs with $q/h^3=1$, $\langle\dot{m}_{\rm b}\rangle$ in the $\gamma=1.6$ case increases by more than one order of magnitude when $\alpha$ increases from $0$ to $0.1$, while $\langle\dot{m}_{\rm b}\rangle$ in the isothermal case only increases by a factor of $\sim 2$ as the baseline is already high.  Furthermore, the runs with $q/h^3=3$ exhibit similar trends, though with smaller amplifications in $\langle\dot{m}_{\rm b}\rangle$.

To better understand the scaling of our results, the accretion rate contributed by viscosity (onto the two binary components) may be approximated as
\begin{align}
  \langle \dot{m}_{\rm b} \rangle_{\rm visc} &\sim 2\times 3\pi \Sigma_{\rm g} \nu = 6\pi \Sigma_{\rm g} \frac{\alpha c_{\rm s}^2}{\Omega} \label{eq:mdot_visc} \\
  \begin{split}\label{eq:mdot_scaling}
    \Rightarrow \frac{\langle \dot{m}_{\rm b} \rangle_{\rm visc}}{\Sigma_{\infty}v_{\rm b} a_{\rm b}} &\sim 6\pi\alpha \left( \frac{\Sigma_{\rm g}}{\Sigma_{\infty}} \right) \left(\frac{c_{\rm s}}{c_{\rm s,\infty}} \right)^2 \left( \frac{c_{\rm s,\infty}}{v_{\rm b}} \right)^2 \left( \frac{\Omega}{\Omega_{\rm b}} \right)^{-1} \\
    &\sim 6\pi \alpha \mathcal{A}_{\Sigma} \mathcal{A}_{c_{\rm s}} \mathcal{A}_{\Omega}^{-1} \left( \frac{q}{h^3} \right)^{2/3} \lambda^{-1},
  \end{split}
\end{align}
where we define $\mathcal{A}_{\Sigma} \equiv \Sigma_{\rm g} / \Sigma_{\infty}$, $\mathcal{A}_{c_{\rm s}} \equiv (c_{\rm s} / c_{\rm s,\infty})^2$, and $\mathcal{A}_{\Omega} \equiv \Omega / \Omega_{\rm b}$.  Taking the case with $q/h^3=1$, $\gamma=1.6$, $\alpha=0.1$ as an example to assess the above equation --- at the radius of $0.25 a_{\rm b}$ around each binary component, $\mathcal{A}_{\Sigma}\mathcal{A}_{c_{\rm s}} \sim 10^{1.5}$ (by eye from Fig. \ref{fig:snap_runII_qth1}) and $\mathcal{A}_{\Omega} \sim 5.76$, we obtain $\langle \dot{m}_{\rm b} \rangle_{\rm visc} \sim 2.1\ \Sigma_{\infty}v_{\rm b} a_{\rm b}$, roughly consistent with the increase of the time-averaged accretion rate between $\alpha=0.0$ and $\alpha=0.1$: $\Delta \langle\dot{m}_{\rm b}\rangle = \mbox{{2.1}} \ \Sigma_{\infty}v_{\rm b} a_{\rm b}$.  \rlr{The same applies to the accretion rate increment when $\alpha$ increases from $0.0$ to $0.01$.}

The scaling laws in Eq. \ref{eq:mdot_scaling} indicate that the accretion rate contributed by viscosity decreases with $q/h^3$ and with decreasing $\gamma$ (due to the decrease in $\mathcal{A}_{\Sigma}\mathcal{A}_{c_{\rm s}}$), consistent with our findings from Fig. \ref{fig:runII_trends_alpha}.  Specifically, for large $\gamma$ (far from isothermal), $\mathcal{A}_{\Sigma} \gtrsim 1$ and $\mathcal{A}_{c_{\rm s}} \gg 1$, whereas for small $\gamma$ (nearly isothermal), $\mathcal{A}_{\Sigma} \gg 1$ and $\mathcal{A}_{c_{\rm s}} \to 1$.  Thus, the product $\mathcal{A}_{\Sigma}\mathcal{A}_{c_{\rm s}}$ moderately decreases with decreasing $\gamma$.  Following the same scaling in Eq. \ref{eq:mdot_scaling}, we can further infer that the viscosity-driven $\dot{m}_{\rm b}$ would decrease with $\lambda$.

Besides accretion, Figs. \ref{fig:dot_mL_runII_qth1} and \ref{fig:runII_trends_alpha} show that viscosity also increases the total torque on the binary, $\langle\dot{L}_{\rm b}\rangle$, with trends similar to the change in accretion rate.  Previous work have focused on how the change in persistent flow structure near the binary (e.g., CSDs, trailing tails) would affect the gravitational torque onto the binary.  Our \citetalias{Li2023} found that the torque associated with accretion is often comparable to the gravitational torque and is thus essential in determining the binary orbital evolution.  In this work, Fig. \ref{fig:runII_dotLs_trends_alpha} shows that the accretion torque is able to dominate the total torque at high viscosities and may solely determine the binary orbital evolution.  
\rlr{Particularly, the second term of the accretion torque (see Eq. \ref{eq:dotL_acc}), $\left\langle \dot{\mu}_{\rm b} (\bm{r}_{\rm b} \times \dot{\bm{r}}_{\rm b}) \right\rangle$, is always positive, increases with $\alpha$, and becomes increasingly dominant at high viscosities.}
\rlr{Meanwhile, the torques associated with pressure and viscosity are always much smaller than the other two leading components and are therefore mostly negligible.}

Indeed, the new behaviour of the binary torque $\langle \dot{L}_{\rm b} \rangle$ can be understood as the result of the viscosity-boosted accretion.
Because the background gas enters the binary Hill sphere at nearly zero velocity (i.e., around the stagnant points where the flow diverges into the horseshoe flow and the shear flow) in the rotating frame, it possesses positive angular momentum relative to the binary in the inertial frame \citep{Tanigawa2002}.  A significant fraction of this angular momentum may be lost when the gas moves inwards (either via shocks or viscosity), but the remaining angular momentum could dominate $\langle\dot{L}_{\rm b}\rangle$ when $\langle\dot{m}_{\rm b}\rangle$ \rlr{(or $\langle\dot{\mu}_{\rm b}\rangle$)} is large enough.  
In other words, viscosity boosts accretion and $\langle\dot{L}_{\rm acc}\rangle$, while the moderate changes in flow structure do not significantly alter $\langle\dot{L}_{\rm grav}\rangle$.  In the cases with $\gamma=1.6$ in Fig. \ref{fig:runII_dotLs_trends_alpha}, $\langle\dot{L}_{\rm grav}\rangle$ does change somewhat more as the CSDs vanish towards higher viscosities, but it is still nowhere near the magnitude of $\langle\dot{L}_{\rm acc}\rangle$ when $\alpha=0.1$.  

Since viscosity increases both $\langle\dot{m}_{\rm b}\rangle$ and $\langle\dot{L}_{\rm b}\rangle$, the binary migration rate \rlr{(for equal-mass circular binaries)}
\begin{equation} \label{eq:adota}
  \frac{\langle\dot{a}_{\rm b}\rangle}{a_{\rm b}} = 8 \left( \frac{\ell_0}{\Omega_{\rm b} a_{\rm b}^2} - \frac{3}{8}\right) \frac{\langle\dot{m}_{\rm b}\rangle}{m_{\rm b}}
\end{equation}
\rlr{may be positive or negative depending on the accretion eigenvalue $\ell_0 \equiv \langle\dot{L}_{\rm b}\rangle / \langle\dot{m}_{\rm b}\rangle$.  Fig. \ref{fig:runII_trends_alpha} shows that $\langle\dot{a}_{\rm b}\rangle / a_{\rm b}$ first increases and then decreases with $\alpha$ in non-isothermal cases, and monotonically decreases with $\alpha$ in isothermal cases.  This finding suggests that $\ell_0$ mostly decreases with $\alpha$ as the outward transport of angular momentum becomes more efficient, indicating that it is generally easier to harden a binary in viscous discs.  Specifically, all cases produce inspiral binaries except the isothermal cases with $q/h^3=3$, where the massive CSDs provide non-negligible positive gravitational torques (see also \citetalias{Li2023}).}

\rlr{Furthermore, Table \ref{tab:runs-visc} shows that the time-averaged spin torque can be expressed as}
\begin{equation}
  \langle \dot{S}_i \rangle \equiv \mathcal{A}_{\dot{S}_i} \langle \dot{m}_i \rangle \sqrt{G m_i r_{\rm e}}, \label{eq:dotS_i}
\end{equation}
\rlr{where $\mathcal{A}_{\dot{S}_i}$ is always close to unity in the cases with prograde binaries and with persistent CSDs, indicating that those CSDs modelled in our simulations have reached a steady-state (see Fig. \ref{fig:dot_S_runII_qth3_ga1.1_a0.1}), regardless of the physical parameters.  This finding also validates the use of Eq. \ref{eq:mdot_visc} for estimating accretion rate.  For the two cases without persistent CSDs (i.e., with $\gamma=1.6$ and $\alpha=0.1$), $\mathcal{A}_{\dot{S}_i}$ deviates from unity as expected.}

\subsection{Retrograde and Eccentric Binaries}
\label{subsec:retrograde_eccentric}

In \citetalias{Li2022}, we studied the orbital evolution of retrograde circular binaries and prograde eccentric binaries in inviscid discs.  In this section, we conduct two additional experiments with $(q/h^3, \gamma, \alpha)=(3, 1.1, 0.1)$ to investigate how these different types of binary orbits are influenced by the viscous discs (see Table \ref{tab:runs-visc}).

Our results for the retrograde binary and eccentric binary are in line with our findings in \citetalias{Li2022}.  The retrograde circular binary does not have persistent CSDs or prominent circumbinary structures as the two components orbit each other rapidly and constantly disrupt the spirals/shocks excited by the other component.  Consequently, the retrograde binary accretes directly from surrounding flows, at a rate about two times its prograde counterpart.  Moreover, the accreted gas mostly brings angular momentum that torques against the retrograde orbit.  Such a torque dominates the total torque onto the binary and leads to fast orbital decay:
\begin{equation}
  \frac{\langle\dot{a}_{\rm b}\rangle}{a_{\rm b}} \simeq -35.90 \frac{\Sigma_\infty v_{\rm b} a_{\rm b}}{m_{\rm b}} \simeq -9.65 \frac{\langle\dot{m}_{\rm b}\rangle}{m_{\rm b}},
\end{equation}
where the value of $\langle\dot{a}_{\rm b}\rangle / a_{\rm b}$ in units of $\langle\dot{m}_{\rm b}\rangle / m_{\rm b}$ is quite similar to those obtained in inviscid runs (see Section 3.3 in \citetalias{Li2022}).

\begin{figure*}
  \centering
  \includegraphics[width=\linewidth]{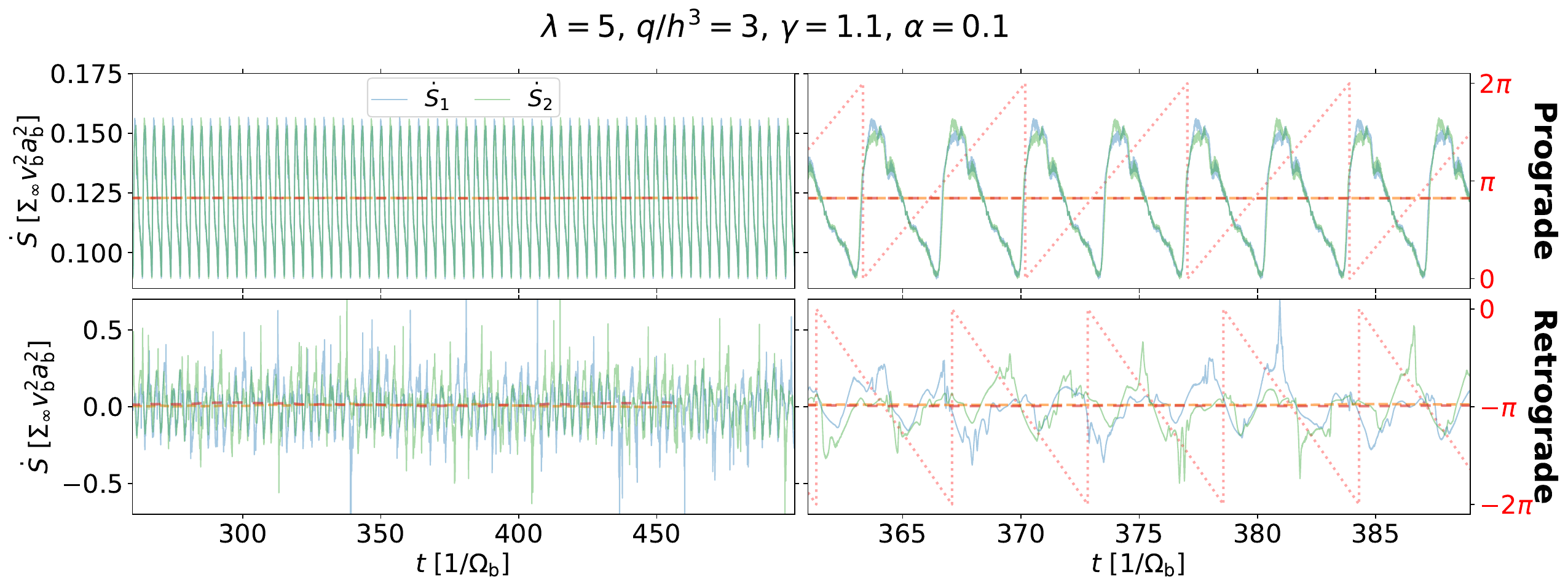}
  \caption{Similar to Fig. \ref{fig:dot_mL_runII_qth1} but compare the spin torques $\dot{\bm{S}}_i$ in runs with prograde (\textit{upper}) and retrograde (\textit{lower}) binaries, both with $\lambda=5$, $q/h^3=3$, $\gamma=1.1$, and $\alpha=0.1$.  
  \label{fig:dot_S_runII_qth3_ga1.1_a0.1}}
\end{figure*}

\rlr{Fig. \ref{fig:dot_S_runII_qth3_ga1.1_a0.1} further compares the time series of the accreted spin torques onto the binary components in prograde and retrograde binaries.  Unlike the well-behaved periodic variations shown in the evolution of $\dot{S}_i$ in the prograde binary, the spin torques in the retrograde binary case exhibit much larger and stochastic variations, again due to the lack of persistent CSDs.  Comparing to the finite $\langle \dot{S}_i \rangle$ obtained by the prograde binary (see Eq. \ref{eq:dotS_i}), the time-averaged spin torques onto the retrograde binary are close to $0$, suggesting that it might be inefficient to align the spins of components in retrograde binaries with the disc gas through accretion.}

The prograde eccentric binary with $e_{\rm b}=0.5$ accretes moderately faster than its circular counterpart and thus receives a more positive total torque, in good agreement with the trends identified in \citetalias{Li2022} (see Section 3.2 in \texttt{Run II-e$_{\mathtt{b}}$}).  The higher accretion rate is due to the fact that binary components can dive farther into the shear flow, where more materials are available for accretion.  Since the circular counterpart is already an outspiral binary, this eccentric binary experiences even faster orbital expansion.  Despite being driven to expand, it still has a short circularization time-scale, where $\langle \dot{e}_{\rm b}^2 \rangle = -0.83 \langle\dot{m}_{\rm b}\rangle/m_{\rm b}$.

\begingroup 
\setlength{\medmuskip}{0mu} 
\setlength\tabcolsep{4pt} 
\setcellgapes{3pt} 
\begin{table}
  \nomakegapedcells
  \caption{Results for the simulation survey with varying sink radius and $\alpha=0.1$} \label{tab:runs-visc01}
  \makegapedcells 
  \linespread{1.025} 
  \begin{tabular}{ccc|rrrr}
    \hline
    $q/h^3$
    & $\gamma$
    & $r_{\rm s}$
    & \makecell[c]{$\langle\dot{m}_{\rm b}\rangle$}
    & \makecell[c]{$\langle\dot{L}_{\rm b}\rangle$}
    & \makecell[c]{$\displaystyle \frac{\langle\dot{a}_{\rm b}\rangle}{a_{\rm b}}$}
    & \makecell[c]{$\displaystyle \frac{ \left\langle\dot{S}_i \right\rangle }{ \langle\dot{m}_i \rangle }$}
    \\
    
    &
    & [$a_{\rm b}$]
    & \makecell[c]{\footnotesize $\displaystyle \left[ \Sigma_\infty v_{\rm b} a_{\rm b} \right]$}
    & \makecell[c]{\footnotesize $\displaystyle \left[ \Sigma_\infty v_{\rm b}^2 a_{\rm b}^2 \right]$}
    & \makecell[c]{\small $\displaystyle \left[ \frac{\Sigma_\infty a_{\rm b} v_{\rm b}}{m_{\rm b}} \right]$}
    & \makecell[c]{\small $\displaystyle \left[ \sqrt{G m_i r_{\rm e}} \right]$}
    \\
    (1)
    & (2)
    & (3)
    & \makecell[c]{(4)}
    & \makecell[c]{(5)}
    & \makecell[c]{(6)}
    & \makecell[c]{(7)}
    \\
    \hline\hline
       &      & $0.02$ &$ 2.22$ &$ 0.55$ &$-2.26$ &$0.82$ \\
       & $1.6$& $0.04$ &$ 2.28$ &$ 0.60$ &$-2.05$ &$0.78$ \\
       &      & $0.08$ &$ 2.37$ &$ 0.67$ &$-1.72$ &$0.74$ \\ \cline{2-7}
       &      & $0.02$ &$ 1.69$ &$ 0.51$ &$-0.95$ &$1.00$ \\
    $1$& $1.1$& $0.04$ &$ 1.82$ &$ 0.56$ &$-0.98$ &$1.00$ \\
       &      & $0.08$ &$ 1.97$ &$ 0.62$ &$-0.93$ &$0.97$ \\ \cline{2-7}
       &      & $0.02$ &$ 1.59$ &$ 0.53$ &$-0.54$ &$1.02$ \\
       &   $1$& $0.04$ &$ 1.79$ &$ 0.55$ &$-0.94$ &$1.03$ \\
       &      & $0.08$ &$ 2.01$ &$ 0.71$ &$-0.34$ &$1.00$ \\
    \Xhline{3\arrayrulewidth}
       &      & $0.02$ &$ 1.42$ &$ 0.44$ &$-0.75$ &$0.81$ \\
       & $1.6$& $0.04$ &$ 1.52$ &$ 0.51$ &$-0.50$ &$0.80$ \\
       &      & $0.08$ &$ 1.63$ &$ 0.55$ &$-0.51$ &$0.78$ \\ \cline{2-7}
       &      & $0.02$ &$ 1.51$ &$ 0.42$ &$-1.20$ &$1.00$ \\
    $3$& $1.1$& $0.04$ &$ 1.63$ &$ 0.52$ &$-0.77$ &$1.00$ \\
       &      & $0.08$ &$ 1.87$ &$ 0.71$ &$ 0.04$ &$0.97$ \\ \cline{2-7}
       &      & $0.02$ &$ 1.93$ &$ 0.87$ &$ 1.14$ &$1.03$ \\
       &   $1$& $0.04$ &$ 2.14$ &$ 0.97$ &$ 1.32$ &$1.05$ \\
       &      & $0.08$ &$ 2.34$ &$ 1.09$ &$ 1.69$ &$1.04$ \\
    \hline
  \end{tabular} \\
  \begin{flushleft}
    {\large N}OTE -- The columns are the same as those in Table \ref{tab:runs-visc} except the third one, which lists the sink radius for in each run. 
  \end{flushleft}
\end{table}
\endgroup

\begingroup 
\setlength{\medmuskip}{0mu} 
\setlength\tabcolsep{4pt} 
\setcellgapes{3pt} 
\begin{table}
  \nomakegapedcells
  \caption{Results for the simulation survey with varying sink radius and $\alpha=0.0$} \label{tab:runs-novisc}
  \makegapedcells 
  \linespread{1.025} 
  \begin{tabular}{ccc|rrrr}
    \hline
    $q/h^3$
    & $\gamma$
    & $r_{\rm s}$
    & \makecell[c]{$\langle\dot{m}_{\rm b}\rangle$}
    & \makecell[c]{$\langle\dot{L}_{\rm b}\rangle$}
    & \makecell[c]{$\displaystyle \frac{\langle\dot{a}_{\rm b}\rangle}{a_{\rm b}}$}
    & \makecell[c]{$\displaystyle \frac{ \left\langle\dot{S}_i \right\rangle }{ \langle\dot{m}_i \rangle }$}
    \\
    
    &
    & [$a_{\rm b}$]
    & \makecell[c]{\footnotesize $\displaystyle \left[ \Sigma_\infty v_{\rm b} a_{\rm b} \right]$}
    & \makecell[c]{\footnotesize $\displaystyle \left[ \Sigma_\infty v_{\rm b}^2 a_{\rm b}^2 \right]$}
    & \makecell[c]{\small $\displaystyle \left[ \frac{\Sigma_\infty a_{\rm b} v_{\rm b}}{m_{\rm b}} \right]$}
    & \makecell[c]{\small $\displaystyle \left[ \sqrt{G m_i r_{\rm e}} \right]$}
    \\
    (1)
    & (2)
    & (3)
    & \makecell[c]{(4)}
    & \makecell[c]{(5)}
    & \makecell[c]{(6)}
    & \makecell[c]{(7)}
    \\
    \hline\hline
       &      & $0.02$ &$ 0.12$ &$-0.16$ &$-1.59$ &$1.03$ \\
       & $1.6$& $0.04$ &$ 0.18$ &$-0.08$ &$-1.18$ &$1.08$ \\
       &      & $0.08$ &$ 0.37$ &$ 0.07$ &$-0.55$ &$1.11$ \\ \cline{2-7}
       &      & $0.02$ &$ 0.50$ &$ 0.03$ &$-1.28$ &$1.00$ \\
    $1$& $1.1$& $0.04$ &$ 0.66$ &$ 0.13$ &$-0.97$ &$1.06$ \\
       &      & $0.08$ &$ 0.86$ &$ 0.25$ &$-0.55$ &$1.05$ \\ \cline{2-7}
       &      & $0.02$ &$ 0.85$ &$ 0.35$ &$ 0.23$ &$1.03$ \\
       &   $1$& $0.04$ &$ 0.97$ &$ 0.35$ &$-0.12$ &$1.06$ \\
       &      & $0.08$ &$ 1.15$ &$ 0.48$ &$ 0.38$ &$1.04$ \\
    \Xhline{3\arrayrulewidth}
       &      & $0.02$ &$ 0.13$ &$-0.05$ &$-0.77$ &$1.07$ \\
       & $1.6$& $0.04$ &$ 0.18$ &$-0.04$ &$-0.86$ &$1.10$ \\
       &      & $0.08$ &$ 0.44$ &$ 0.10$ &$-0.50$ &$1.10$ \\ \cline{2-7}
       &      & $0.02$ &$ 0.83$ &$ 0.12$ &$-1.53$ &$1.01$ \\
    $3$& $1.1$& $0.04$ &$ 1.10$ &$ 0.38$ &$-0.27$ &$1.05$ \\
       &      & $0.08$ &$ 1.56$ &$ 0.72$ &$ 1.07$ &$1.05$ \\ \cline{2-7}
       &      & $0.02$ &$ 1.74$ &$ 0.93$ &$ 2.24$ &$1.04$ \\
       &   $1$& $0.04$ &$ 1.84$ &$ 0.89$ &$ 1.64$ &$1.06$ \\
       &      & $0.08$ &$ 2.06$ &$ 1.02$ &$ 1.99$ &$1.06$ \\
    \hline
  \end{tabular} \\
  \begin{flushleft}
    {\large N}OTE -- The columns are the same as those in Table \ref{tab:runs-visc01}. 
  \end{flushleft}
\end{table}
\endgroup

\begin{figure*}
  \centering
  \includegraphics[width=\linewidth]{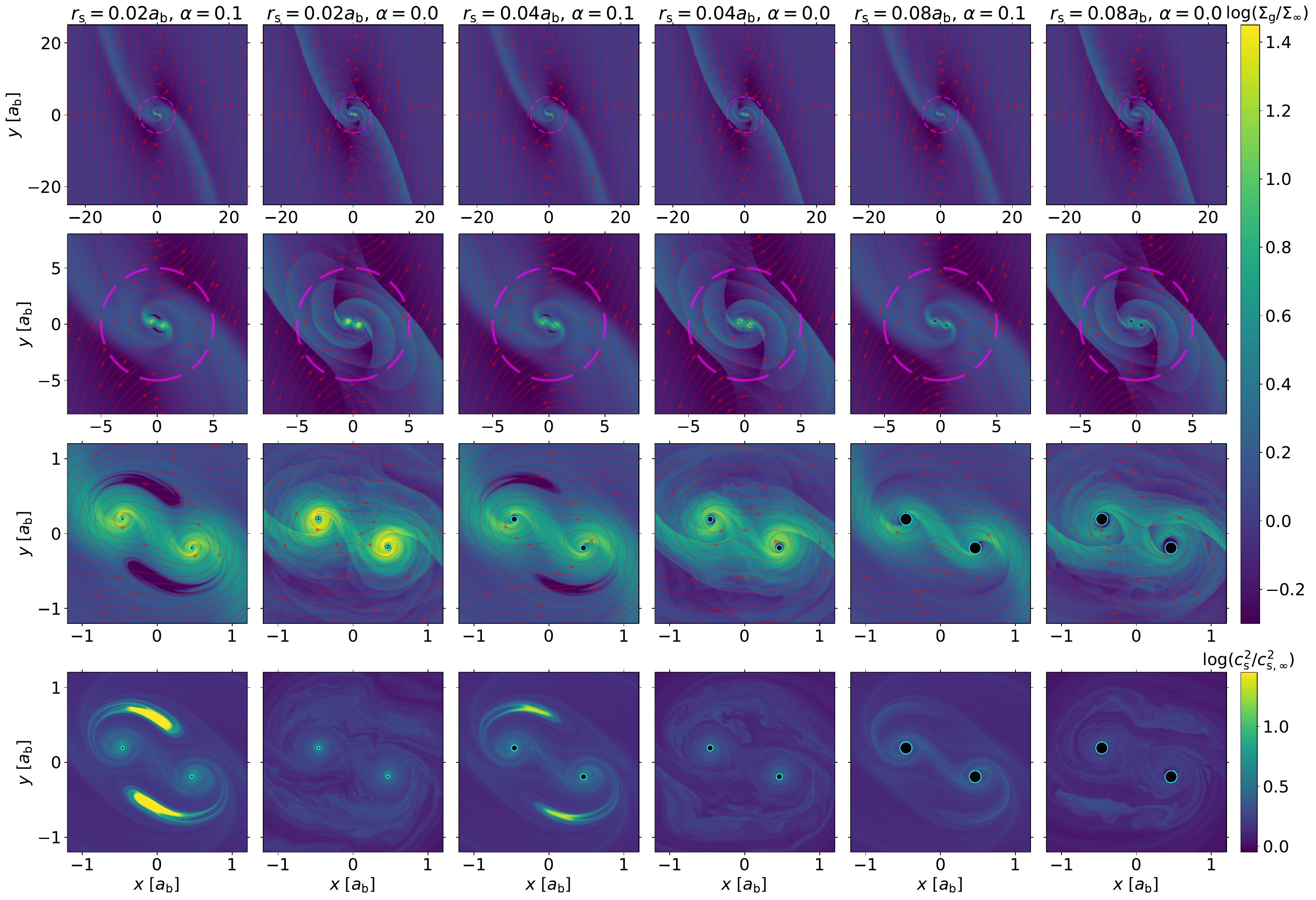}
  \caption{Similar to Fig. \ref{fig:snap_runII_qth1} but for runs with $\lambda=5$, $q/h^3=1$, $\gamma=1.1$, and  (from \textit{left} to \textit{right}) $r_{\rm s}=0.02 a_{\rm b}$, $0.04 a_{\rm b}$, and $0.08 a_{\rm b}$.
  \label{fig:snap_runII_qth1_ga1.1_rs}}
\end{figure*}

\begin{figure*}
  \centering
  \includegraphics[width=0.495\linewidth]{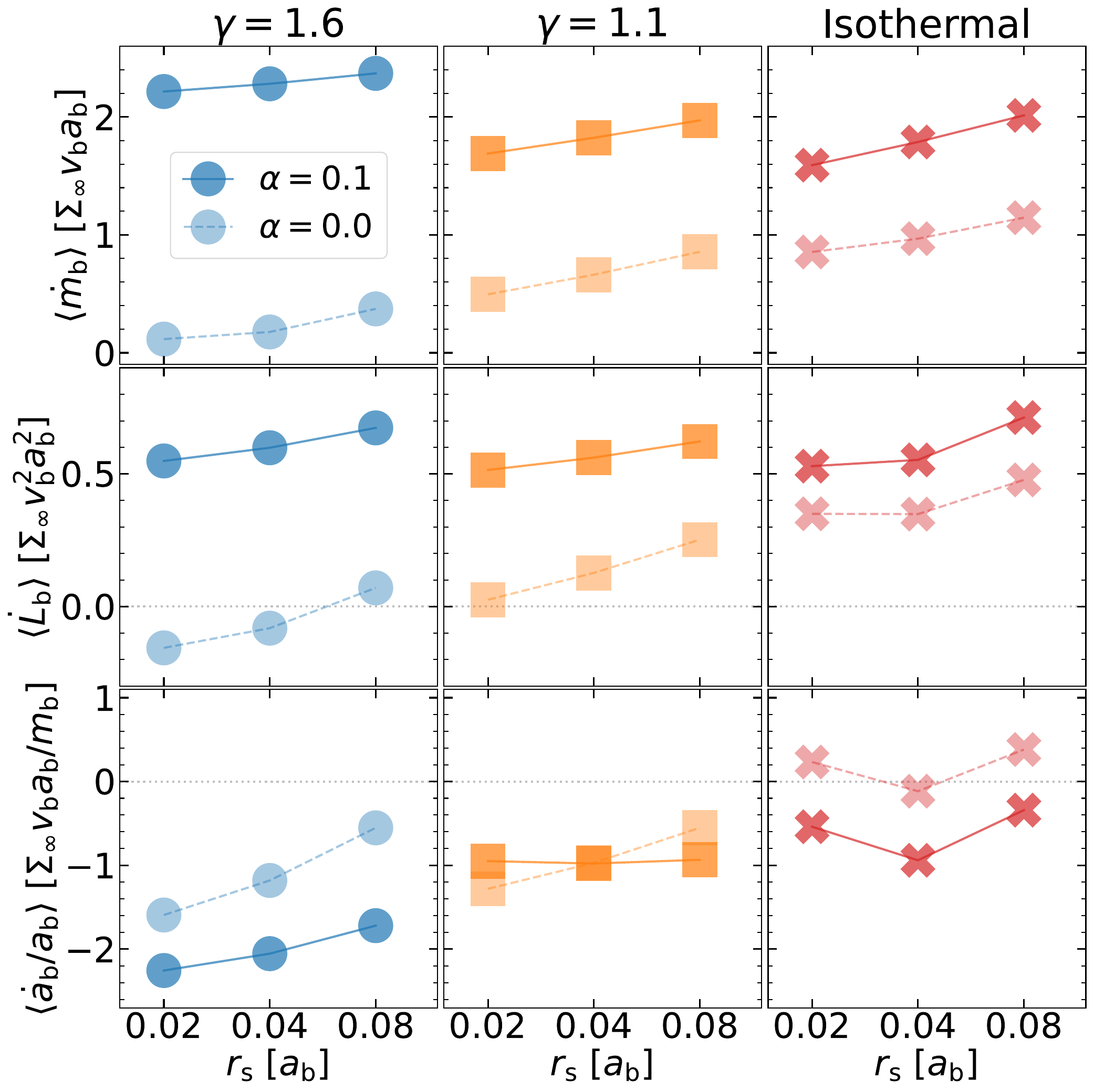}
  \includegraphics[width=0.495\linewidth]{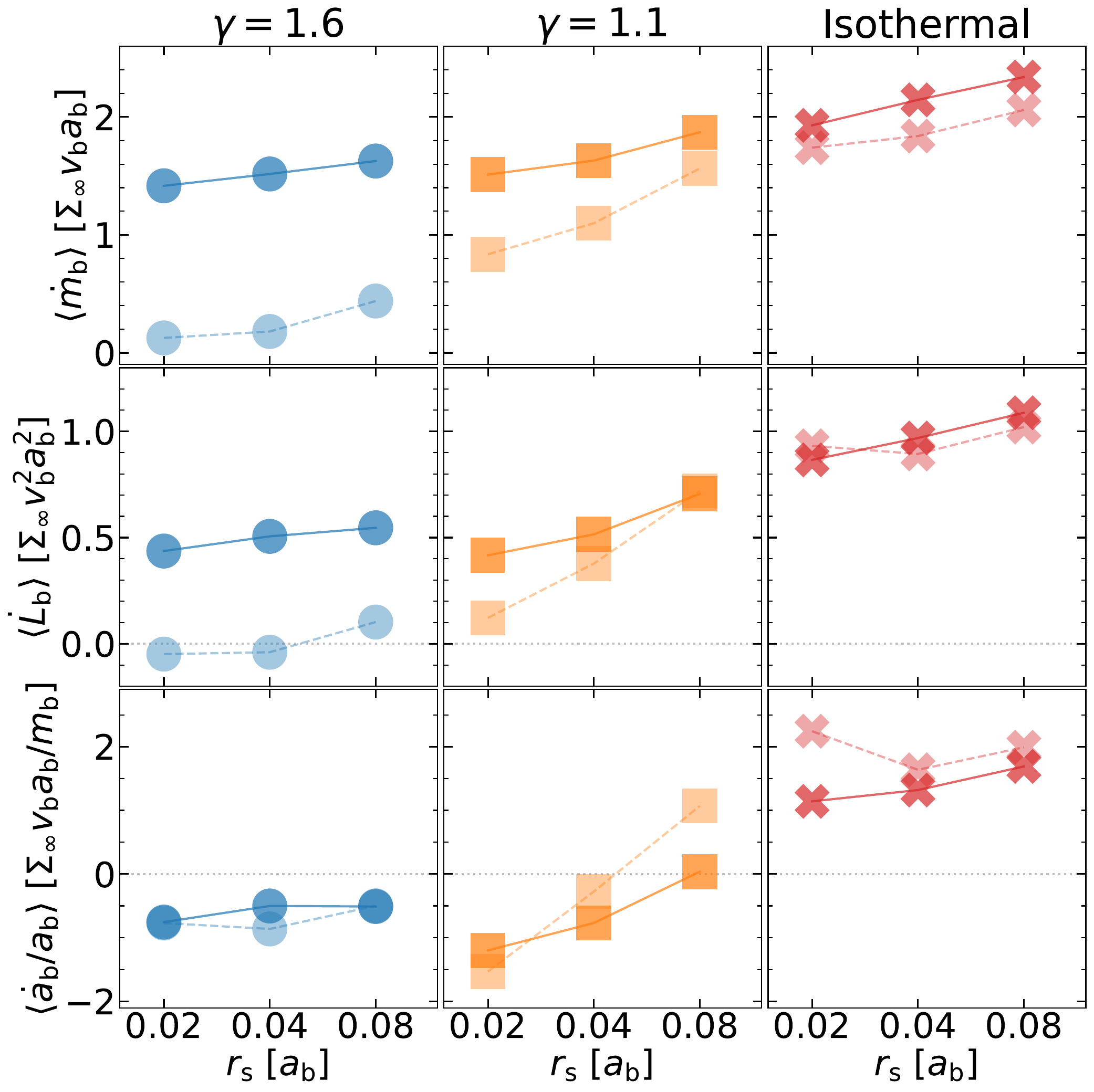}
  \caption{Similar to Fig. \ref{fig:runII_trends_alpha} but showing the results as a function of the sink radius $r_{\rm s}$, for runs with $\alpha=0.1$ (\textit{solid lines}) and with $\alpha=0.0$ (\textit{dashed lines})\rlr{, and for runs with $q/h^3=1$ (\textit{left}) and with $q/h^3=3$ (\textit{right})}.
  \label{fig:runII_trend_rs}}
\end{figure*}

\section{Dependences on Sink Radius}
\label{sec:deps_on_r_s}

Our canonical simulations described in Section \ref{sec:results} assume that each binary component has a sink radius $r_{\rm s}=0.04 a_{\rm b}$.  In this section, we perform experiments on the evolution of embedded binaries with varying sink radius $r_{\rm s}$.  \citetalias{Li2022} explored the effects of $r_{\rm s}$ in inviscid flow with $\gamma=1.6$ and found that the accretion rate and the orbital evolution rate are largely influenced by the choice of $r_{\rm s}$ (see their Section 3.1.5), indicating the need of linking the sink radius to a physically motivated size of the accretor.  However, viscous accretion onto isolated binaries seems to be rather independent of the sink radius since the accretion through a circumbinary disc is regulated by viscosity \citep[e.g.,][]{Munoz2020}.

It is thus of great interest to see if viscosity can reduce the accretion rate dependency
\footnote{\rlr{This dependency focuses on the relative fraction of change in the accretion rate due to the change in $r_{\rm s}$, not the slope.}}
on the sink radius for embedded binaries.  Tables \ref{tab:runs-visc01} and \ref{tab:runs-novisc} summarize the parameters and results of our simulation survey on $r_{\rm s}$.  For each $q/h^3$ ($1$ or $3$) and each $\gamma$ ($1.6$, $1.1$, or isothermal), we perform three viscous ($\alpha=0.1$) simulations with $r_{\rm s}=0.02 a_{\rm b}$, $0.04 a_{\rm b}$, and $0.08 a_{\rm b}$, respectively, which covers half and double the fiducial value of $r_{\rm s}$ (see Table \ref{tab:runs-visc01}).  Furthermore, we conduct the same set of $18$ simulations but with inviscid flows ($\alpha=0.0$; see Table \ref{tab:runs-novisc}) as a control group that enables us to better understand the effect of viscosity.

Fig. \ref{fig:snap_runII_qth1_ga1.1_rs} compares the flow structure between different sink radii for both viscous and inviscid runs in the final quasi-steady state.  Similar to our findings in \citetalias{Li2022}, we find that accretors with smaller $r_{\rm s}$ have less truncated inner cavities and hence more massive CSDs, regardless of viscosity.  Also, the pair of crescent voids become more prominent when $r_{\rm s}=0.02 a_{\rm b}$ but almost disappear when $r_{\rm s}=0.08 a_{\rm b}$, suggesting that the sink radius may fundamentally alter the flow morphology in the vicinity and a large $r_{\rm s}$ could suppress not only the CSDs but also other features in the circumbinary flows.  In addition, we find that the CSDs in the viscous runs are always less massive than their inviscid counterparts, consistent with the results shown in Fig. \ref{fig:snap_runII_qth1}.  

Fig. \ref{fig:runII_trend_rs} presents the secular orbital evolution results as a function of $r_{\rm s}$ for our survey with different $q/h^3$, $\gamma$, and $\alpha$.  For the inviscid runs, we find that the accretion rate increases with decreasing $\gamma$ and with increasing $q/h^3$, consistent with our findings in \citetalias{Li2023}.  In addition, $\langle \dot{m}_{\rm b} \rangle$ in the inviscid runs monotonically increases with $r_{\rm s}$, consistent with our findings in \citetalias{Li2022}.  We find that these trends remain true for all the viscous runs since it is always easier for streamlines to intersect a larger accretor and get accreted.

We further find that the accretion rate increment from an inviscid run to its viscous counterpart, $\Delta \langle\dot{m}_{\rm b}\rangle$, does not simply scale with $r_{\rm s}$, \textit{but} follows the scaling laws discussed in Section \ref{subsec:orbital_evolution} (see Eq. \ref{eq:mdot_scaling}), i.e., $\Delta \langle\dot{m}_{\rm b}\rangle$ increases with increasing $\gamma$ and decreasing $q/h^3$.  Therefore, given that $\gamma$ and $q/h^3$ affect $\langle \dot{m}_{\rm b} \rangle$ in the inviscid runs and $\Delta \langle\dot{m}_{\rm b}\rangle$ in the viscous runs in different ways, the resulting $\langle \dot{m}_{\rm b} \rangle$ in viscous runs decreases with decreasing $\gamma$ when $q/h^3=1$ but increases with decreasing $\gamma$ at $q/h^3=3$, leading to a mixed $q/h^3$ dependency of $\langle \dot{m}_{\rm b} \rangle$ when $\gamma$ is fixed.

Compared to the inviscid runs with $\gamma=1.6$ where \citetalias{Li2022} found $\langle \dot{m}_{\rm b} \rangle \propto r_{\rm s}$, the other simulations summarized in Tables \ref{tab:runs-visc01} and \ref{tab:runs-novisc} show that this linear scaling with $r_{\rm s}$ becomes weaker for a more isothermal-like EOS or for the viscous cases.  That said, as $\langle \dot{m}_{\rm b} \rangle$ increases with $r_{\rm s}$, the total torque $\langle \dot{L}_{\rm b} \rangle$ also generally increases with $r_{\rm s}$, which makes the $r_{\rm s}$ dependency of the binary migration rate complicated (see Eq. \ref{eq:adota}).  For the viscous runs, Fig. \ref{fig:runII_trend_rs} shows that $\langle\dot{a}_{\rm b}\rangle / a_{\rm b}$ mostly increases with $r_{\rm s}$, suggesting that it is generally more difficult to harden a binary with larger accretor sizes.

\rlr{In addition, Tables \ref{tab:runs-visc01} and \ref{tab:runs-novisc} show that the time-averaged spin torques in all cases (again except those without persistent CSDs, i.e., for $\gamma=1.6$ and $\alpha=0.1$) closely follow Eq. \ref{eq:dotS_i}, with $\mathcal{A}_{\dot{S}_i} \approxeq 1$.  This finding is consistent with results shown in Section \ref{subsec:orbital_evolution} and indicates that our simulations have reached a steady-state, regardless of the numerical parameter $r_{\rm s}$.}

\section{Summary}
\label{sec:summary}

In this follow-up study to our \citetalias{Li2022} and \citetalias{Li2023}, we perform, for the first time, a suite of 2D, local shearing box, viscous hydrodynamical simulations to study the evolution of BBHs embedded in AGN discs.  We focus on equal-mass, prograde, circular binaries.  Our  survey extensively covers the parameter space of three key dimensionless parameters (see Table \ref{tab:runs-visc}), $q/h^3\equiv (m_{\rm b}/M)(R/H_{\rm g})^3 \equiv (R_{\rm H}/H_{\rm g})^3$, $\gamma$ (in the $\gamma$-law EOS), and $\alpha$ (in the $\alpha$-viscosity prescription).  Moreover, two additional experiments are conducted to study retrograde and eccentric binaries.  As in our previous papers, our simulations resolve the viscous accretion flow down to $\lesssim 1\%$ of the Hill radius around each binary component (treated as an absorbing sphere with a fiducial sink radius $r_{\rm s}=0.04 a_{\rm b}$), with the finest cells $\lesssim 0.1\%$ of the Hill radius.  We also consider binaries modelled with half/double the fiducial sink radius in a numerical survey (see Tables \ref{tab:runs-visc01} and \ref{tab:runs-novisc}) to study the $r_{\rm s}$ dependency of the binary evolution.

Our key findings are as follows:
\begin{enumerate}
  \item Viscosity smooths the shocks excited by the binary components (see Fig. \ref{fig:snap_runII_qth1}), makes the CSDs moderately less massive, boosts the accretion rates (see Eq. \ref{eq:mdot_scaling}), increases the total torques on the binaries, and makes them generally \rlr{easier to be hardened, especially at high viscosities} (see Fig. \ref{fig:runII_trends_alpha}).
  \item In the main numerical survey with the fiducial sink radius, all prograde binaries \rlr{contract (except the isothermal cases with $q/h^3=3$).  The time-averaged binary migration rate $\langle \dot{a}_{\rm b} \rangle / a_{\rm b}$ broadly decreases with $\gamma$ and increases with $q/h^3$, regardless of $\alpha$, consistent with our findings in \citetalias{Li2023}.}
  \item Our case studies on retrograde and eccentric binaries in viscous discs produce results expected from our findings in \citetalias{Li2022}.  Specifically, the CSDs do not form around the retrograde binary, and the fast accretion onto the binary results in its rapid orbital decay.  The prograde eccentric binary also accretes faster and thus receives more positive torques that lead to its orbital expansion, but still with significant eccentricity damping.
  \item \rlr{The spin torques onto the components of prograde binaries with persistent CSDs are in good agreement with expectations from accretors with steady-state CSDs (see Eq. \ref{eq:dotS_i} and Fig. \ref{fig:dot_S_runII_qth3_ga1.1_a0.1}).  In contrast, the components in retrograde binaries receive negligible spin torques due to the lack of CSDs, indicating that it might be inefficient to align their spins with the disc gas.}
  \item In the numerical survey on the sink radius, we find that both viscosity and a more isothermal-like EOS can substantially reduce the accretion rate dependency on $r_{\rm s}$ (see Fig. \ref{fig:runII_trend_rs}).  Such a dependency seems to be only important when $\gamma=1.6$ and $\alpha=0.0$ (i.e., in inviscid flows).  Moreover, the accretion rate contributed by viscosity barely depends on $r_{\rm s}$, in line with the physical expectation.  That said, when the EOS is nearly isothermal, a considerable amount of accretion may not be driven by viscosity and the total accretion rate can retain a weak dependence on $r_{\rm s}$.
  \item We find that it is generally more difficult to harden a binary with a larger accretor size (i.e., the sink radius $r_{\rm s}$) in viscous discs.  Still, a small sink radius (e.g., $r_{\rm s} = 0.02 a_{\rm b}$) could lead to binary inspiral, especially when the EOS is non-isothermal ($\gamma>1$).
\end{enumerate}

Both the main numerical survey (Table \ref{tab:runs-visc}) and the additional survey (Tables \ref{tab:runs-visc01} and \ref{tab:runs-novisc}) in this work provide a crucial expansion to the parameter space covered by our prior works \citepalias{Li2022, Li2023}.  In particular, viscosity can qualitatively change the flow dynamics and the secular evolution of the binary.  Admittedly, the isotropic, homogeneous viscosity parameterized by $\alpha$ is idealized.
Certain physical processes like BH feedback, magnetic forces, and radiative transfer may hamper the accretion and make binary hardening easier.  Besides these missing physics, our results are subject to the limitations of the 2D local shearing box approximation.  \rlr{It is also worth conducting comparative studies between our method for modeling accretors and other methods, e.g., timescale/fraction-based gas removal accretion \citep[e.g.,][]{Farris2014} or torque-free sinks \citep[e.g.,][]{Dittmann2021, Dempsey2022}, to better understand the pros and cons of different accretor treatments.}  Ultimately, future 3D simulations with more realistic physical ingredients, both local and global, are required to better model the interactions between BBHs and AGN discs.

\section*{Acknowledgements}
This work has been supported in part by the NSF grant AST-2107796.  Resources supporting this work were provided by the NASA High-End Computing (HEC) Program through the NASA Center for Climate Simulation (NCCS) at Goddard Space Flight Center.  RL thanks Kaitlin Kratter, Hui Li, Diego {Mu{\~n}oz}, Ya-Ping Li, Adam Dempsey, Zoltan Haiman, Yan-Fei Jiang, Paul Duffell, and Barry McKernan for inspiring discussions and useful conversations.

\section*{Data Availability}
The simulation data underlying this paper will be shared on reasonable request to the corresponding author.

\bibliographystyle{mnras}
\bibliography{refs}



\appendix

\section{\rlr{Orbital Evolution with Updated Accretion Rate Evaluation}}
\label{appsec:corrected_mdot}

\rlr{In this section, we present the orbital evolution results from new simulations that are numerically identical to two series of surveys shown in our \citetalias{Li2023} (one with $\lambda=5$ and the other with $q/h^3=1$) but with updated method for accretion rate evaluation (see Eqs. \ref{eq:m_dot_i} and \ref{eq:f_hydro}).}

\begin{figure*}
  \centering
  \includegraphics[width=0.495\linewidth]{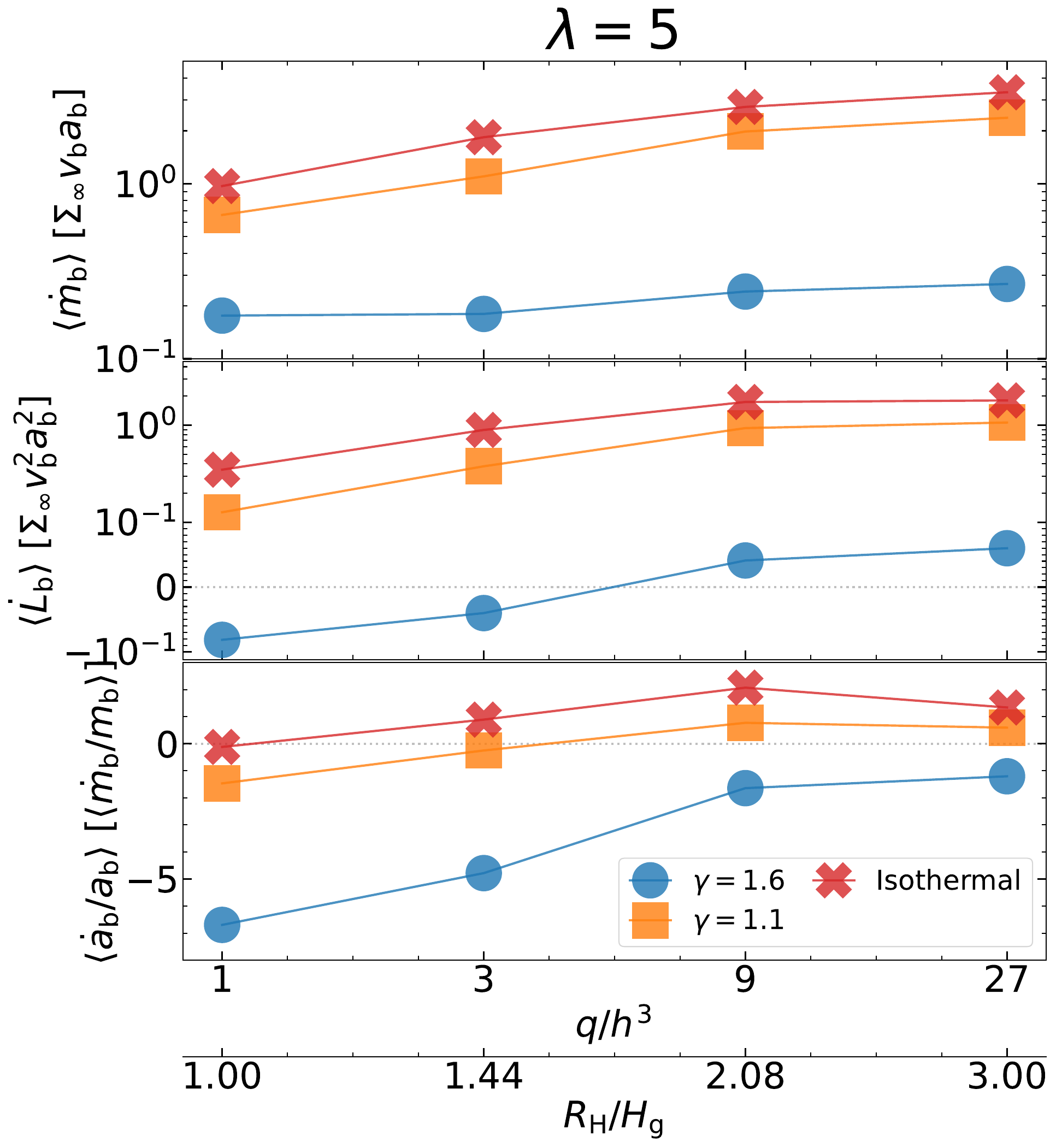}
  \includegraphics[width=0.495\linewidth]{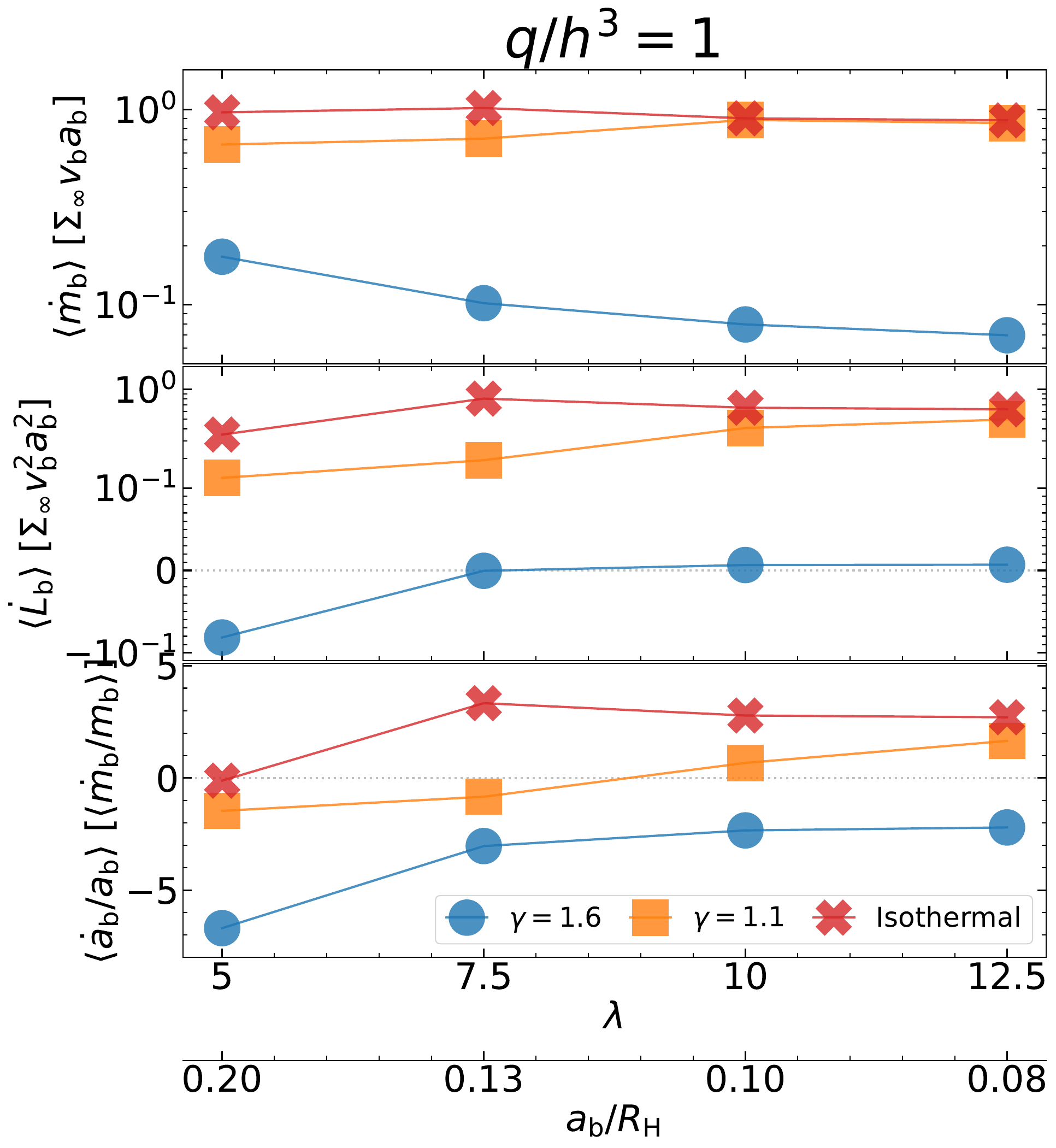}
  \caption{Time-averaged measurements of (from \textit{top} to \textit{bottom}) accretion rate $\langle\dot{m}_{\rm b}\rangle$, total torque $\langle\dot{L}_{\rm b}\rangle$, and binary migration rate $\langle\dot{a}_{\rm b}\rangle/a_{\rm b}$ as a function of $q/h^3$ with $\lambda=5$ (\textit{left}; as a comparison to the left panel of Fig. 4 in \citetalias{Li2023}) or as a function of $\lambda$ with $q/h^3=1$ (\textit{right}; as a comparison to the left panel of Fig. 5 in \citetalias{Li2023}), color-coded by the EOS ($\gamma=1.6$: \textit{blue circle}; $1.1$: \textit{orange square}; isothermal: \textit{red cross}).  
  \label{appfig:runII_qth1_trends}}
\end{figure*}

\rlr{As a comparison to the left panels of Figs. 4 and 5 in \citetalias{Li2023}, Fig. \ref{appfig:runII_qth1_trends} presents the secular orbital evolution results for our fixed-$\lambda$ survey as a function of $q/h^3$, as well as for our fixed-$q/h^3$ survey as a function of $\lambda$, grouped by the EOS.  We omit the cases with $\gamma=1.001$ as they have been shown to produce results in good agreement with the isothermal cases.  Fig. \ref{appfig:runII_qth1_trends} demonstrates that the new method does not affect the conclusions in our previous works, particularly regarding how the orbital evolution of BBHs depends on the three dimensionless parameters: $\gamma$ (in the $\gamma$-law EOS), $q/h^3\equiv (m_{\rm b}/M)(R/H_{\rm g})^3 \equiv (R_{\rm H}/H_{\rm g})^3$ and $\lambda = R_{\rm H}/a_{\rm b}$.}

\bsp    
\label{lastpage}
\end{document}